\documentclass[aps,prd,preprint,floatfix,superscriptaddress]{revtex4}

\usepackage{graphicx}

\def\ra{\rightarrow}

\def\bwt{\begin{widetext}}
\def\ewt{\end{widetext}}
\def\be{\begin{equation}}
\def\ee{\end{equation}}
\def\bea{\begin{eqnarray}}
\def\eea{\end{eqnarray}}
\def\bean{\begin{eqnarray*}}
\def\eean{\end{eqnarray*}}
\def\bary{\begin{array}}
\def\eary{\end{array}}
\def\bit{\begin{itemize}}
\def\eit{\end{itemize}}

\def\ra{\rightarrow}

\begin{document}

\setcounter{page}{0}
\thispagestyle{empty}

\preprint{ANL-HEP-PR-02-022, EFI-02-78, hep-ph/0205342}
\bigskip

\title{Higgs Boson Decay into Hadronic Jets}

\author{Edmond L.\ Berger}
\email[e-mail: ]{berger@anl.gov}
\affiliation{High Energy Physics Division, 
Argonne National Laboratory, Argonne, IL 60439}
\author{Cheng-Wei Chiang}
\email[e-mail: ]{chengwei@hep.uchicago.edu}
\affiliation{High Energy Physics Division, 
Argonne National Laboratory, Argonne, IL 60439}
\affiliation{Enrico Fermi Institute and Department of Physics,
University of Chicago, Chicago, IL 60637}
\author{Jing~Jiang}
\email[e-mail: ]{jiangj@hep.anl.gov}
\affiliation{High Energy Physics Division, 
Argonne National Laboratory, Argonne, IL 60439} 
\author{Tim M.\ P.\ Tait}
\email[e-mail: ]{tait@anl.gov}
\affiliation{High Energy Physics Division, 
Argonne National Laboratory, Argonne, IL 60439}
\author{Carlos E.\ M.\ Wagner}
\email[e-mail: ]{cwagner@hep.anl.gov}
\affiliation{High Energy Physics Division, 
Argonne National Laboratory, Argonne, IL 60439}
\affiliation{Enrico Fermi Institute and Department of Physics,
University of Chicago, Chicago, IL 60637}

\date{\today}

\begin{abstract}
The remarkable agreement of electroweak data with standard model (SM) 
predictions motivates the study of extensions of the SM in which the Higgs 
boson is light and couples in a standard way to the weak gauge bosons.  
Postulated new light particles should have small couplings to the gauge 
bosons. Within this context it is natural to assume that the branching fractions 
of the light SM-like Higgs boson mimic those in the standard model. This assumption 
may be unwarranted, however, if there are non-standard light particles coupled 
weakly to the gauge bosons but strongly to the Higgs field. In particular,
the Higgs boson may effectively decay into hadronic jets, possibly without important 
bottom or charm flavor content. As an example, we present a simple extension of the 
SM, in which the predominant decay of the 
Higgs boson occurs into a pair of light bottom squarks that, in turn, manifest 
themselves as hadronic jets. Discovery of the Higgs boson remains possible 
at an electron-positron linear collider, but prospects at hadron colliders are 
diminished substantially.
\end{abstract}


\maketitle

%%%%%%%%%%%%%%%%%%%%%%%%%%%%%%%%%%%%%%%%

%%%%%%%%%%%%%%%%%%%%%%%%%%%%%%
\section{INTRODUCTION}

In the standard model (SM) of elementary particle interactions, 
breaking of electroweak symmetry is achieved through the Higgs 
mechanism.  The simplest realization is provided
by the introduction of a complex Higgs doublet, which leads to
the presence of a neutral 
CP-even Higgs boson $H^0$ in the physical spectrum. 
This state has not been observed, but a good theoretical
description of the precision electroweak data \cite{Abbaneo:2001ix}
within the SM requires the Higgs boson to be lighter than about 200 
GeV~\cite{lephiggs}. 
Although it can be argued that there are internal inconsistencies in the
data~\cite{Chanowitz} that may demand the presence of new 
physics~\cite{Altarelli:2001wx,wethree}, the SM with a light Higgs boson 
provides a surprisingly
good description of the data. This success has induced an overwhelming
preference for weakly interacting extensions of the SM, 
incorporating a light Higgs boson in a natural way, in comparison with heavy
Higgs boson models in which the effect of the large Higgs boson mass in the
oblique corrections is compensated by new physics 
contributions~\cite{Peskin,Chivukula:2000px,wethreeII}.

Among the possible extensions of the SM, the minimal supersymmetric
standard model (MSSM) has been considered most seriously. The minimal
realization of the Higgs mechanism within supersymmetric extensions
of the standard model requires the presence of two Higgs doublets at
low energies.  In most regions of the 
supersymmetry (SUSY) breaking parameter space, the 
lightest neutral CP-even Higgs particle $h$ resembles the SM Higgs boson 
in many of its properties~\cite{Gunion:1984yn}.
Searches for experimental manifestations of the Higgs states are a central 
motivation for the experimental programs at the Fermilab 
Tevatron and the CERN Large Hadron Collider 
(LHC), with experimental detection techniques guided by 
theoretical expectations about the anticipated properties of these states. 

Within the MSSM, the upper bound on the mass of the lightest Higgs state 
is roughly 135 GeV~\cite{Heinemeyer:1998jw}.  For Higgs boson masses $m_h$ 
between 115 and 135 GeV, the total SM decay width is predicted to grow  
from about 3 to about 6 MeV~\cite{hdecay}.  At $m_h = 120$ GeV, the principal 
decay mode is into a pair of bottom quarks $b \overline{b}$, with about 69\% branching 
fraction; this SM branching fraction drops to about 34\% at $m_h = 140$ GeV 
while branching fractions into weak boson decays increase. In weakly interacting 
extensions of the SM, it is natural to assume that the light SM-like Higgs 
boson state has decay branching ratios similar to those in the SM. This 
expectation may be modified easily under the presence of light particles, weakly 
coupled to the weak gauge bosons, but strongly coupled to the Higgs field. 
The resulting Higgs boson decay properties will depend on the rates for decay 
to these new particles.  For instance, the possible decay of the Higgs boson into 
stable neutral particles, such as neutralinos in the MSSM or neutrinos within models 
of extra dimensions, may lead to a Higgs particle with mainly invisible 
decays~\cite{martinwells}.  Alternatively, the Higgs boson may decay 
predominantly into hadronic jets, without any particular bottom or charm content.  
In this article, we consider scenarios which lead to this latter possibility. 

Direct experimental searches at the CERN Large Electron Positron Collider 
(LEP) place the mass of a SM-like Higgs state, with a significant decay 
branching ratio into bottom ($b$) quarks, above approximately 
115 GeV~\cite{lephiggs}. An alternative analysis, based only on the assumption
of Higgs boson decay into hadronic jets, without $b$-tagging, leads to a bound 
of about 113 GeV~\cite{flavorind}. In this article, we explore in detail the 
possible detection of such a Higgs boson at future hadron and lepton 
colliders. 

A Higgs boson with a dominant effective decay branching ratio into
jets may be obtained within the MSSM, under the assumption
of the presence of light bottom squarks in the spectrum.
The possible existence of bottom squarks $\tilde{b}$ with 
low masses is advanced in several recent papers~\cite{Carena:2000ka,
Berger:2000mp}. 
Bottom squarks are the spin-0, charge -1/3, and color triplet 
supersymmetric partners of bottom quarks.  Interestingly, very small
$\tilde{b}$ masses on the order of 10 GeV may be compatible with
existing measurements~\cite{Carena:2000ka,Berger:2000mp,Berger:2001jb,
Berger:2002kc,Dedes:2000nv,Nierste:2000ez,Becher:2001zb,Cao:2001rz,
Leibovich:2002qp,Nappi:1981ft,Behrend:1986md,Savinov:2000jm}.
Within SUSY theories, a light bottom 
squark is obtained most readily for large values of 
$\tan \beta$~\cite{Carena:2000ka}, the ratio of neutral Higgs field vacuum 
expectation values, and we work in this limit.  
Moderate to high values of $\tan \beta$ are further motivated by the fact that 
experiments at LEP II did not find conclusive evidence of the light SUSY 
Higgs boson; such values are favored in order that the predicted mass of 
the Higgs boson remain above the value excluded 
experimentally~\cite{Heinemeyer:1998jw}.  We 
restrict $\tan \beta \lesssim 50$, as for larger values the bottom quark 
couplings to some of the Higgs particles can be strong enough that 
perturbation theory breaks down.  Within the light bottom squark scenario,
the dominant Higgs decay is into a pair of bottom 
squarks $\tilde{b} {\tilde{b}}^*$~\cite{Carena:2000ka} that, in turn, 
manifest themselves as jets of hadrons.   The total width is predicted to increase 
by a factor of ten to several hundred, depending upon the value of 
$\tan \beta$.  Since the couplings to SM particles remain approximately 
unchanged, the upshot is that branching fractions into conventional decay 
modes ($b {\overline b}$, $W W^*$, $Z Z^*$, $gg$, $\tau \overline{\tau}$, $\gamma \gamma$, 
$\dots$) are all reduced by a corresponding factor.  

In order to fix the framework, we concentrate for the most part on the
particular example of a light bottom squark. While details of our approach
depend on the existence of low mass $\tilde{b}$'s, the principal conclusions
are illustrative of the challenges to be faced if the dominant decays of a
light Higgs state, with $m_H < 135$ GeV, are into hadronic jets without
specific flavor tags. In Sec.~II, we summarize salient aspects of the
phenomenology of bottom squarks, including constraints on their couplings, and
we review available experimental bounds.  In Sec.~III, we compute the Higgs
boson width for decay into a pair of bottom squarks as well as the influence of
bottom squarks in loop processes that describe decay into other final states.
The decay width into the gluon-gluon $gg$ final state is
enhanced as is the partonic production cross section $g g \rightarrow h$.  We
show that decay to a pair of bottom squarks is by far the dominant decay mode
of the Higgs boson for large values of $\tan\beta$.  Since the SM decay
couplings are essentially unaffected, the total decay width of the Higgs boson
is increased and the branching fractions into the SM decay modes are decreased
accordingly.  Except for the gluon fusion process, the Higgs boson production 
rates are not enhanced in hadron
collisions and in electron-positron annihilation processes.  As we discuss in
Sec.~IV, dominant decay into bottom squarks that materialize as hadronic jets
makes it much more difficult, if not impossible, to discover the Higgs boson at
a hadron collider.  The possibilities at an electron-positron linear colliders
are examined in Sec.~V where we demonstrate that it remains possible to
discover the Higgs boson and to measure its mass and several of its coupling
strengths.

\section{Low Mass Bottom Squarks}

Light bottom squarks are discussed in Ref.~\cite{Berger:2000mp} in the context
of an explanation for the large bottom quark production cross section at hadron
colliders.  In that work, a light gluino is also postulated with 100\%
branching fraction into a bottom quark and a bottom squark.  In this discussion
of Higgs boson decay, we need not assume a light gluino since there is a direct
coupling of the Higgs boson to a pair of bottom squarks.  The bottom squark is
the LSP, the SUSY particle with lowest mass.  It may decay promptly through
baryon-number and $R$-parity violating interactions into light
quarks~\cite{Berger:2000zk}, or it could be stable on collider time
scales~\cite{cosmological}.  The least model-dependent statement one can make
is that at high energies the $\tilde{b}$ is likely to manifest itself 
experimentally as a jet of hadrons in the detector.  If we introduce 
somewhat more model-dependent assumptions about decay modes of the 
$\tilde{b}$, identification of the Higgs boson at hadron colliders could 
be facilitated if jets containing charm and/or leading baryons can be 
identified cleanly.

The lighter bottom squark is a mixture of the scalar partners of the
left- and right-chiral bottom quarks.  After SUSY breaking and electroweak
symmetry breaking, the mass matrix for bottom squarks in the weak eigenstate
basis is 
\bea
\label{eq:massmatrix}
%\left( \tilde{b}_L^* \: \tilde{b}_R^* \right)
\left( 
\begin{array}{cc}
m^2_{\tilde Q} + m_b^2 + D_L & m_b \left[ A_b - \mu \tan \beta \right] \\
m_b \left[ A_b - \mu \tan \beta \right] &   m^2_{\tilde b} + m_b^2 + D_R
\end{array}
\right) ,
\eea
where $m^2_{\tilde Q}$ and $m^2_{\tilde b}$ are the SUSY-breaking masses
for the third family squark doublet and down-type singlet, respectively,
$A_b$ is the SUSY-breaking interaction term for the Higgs boson and bottom 
squarks, $m_b$ is the bottom quark mass, $\mu$ is the Higgsino mass parameter,
and $D_L$ and $D_R$ are the $D$-terms for the bottom quark sector, given by
$m_Z^2 \cos 2 \beta (-1/2 + \sin^2 \theta_W /3)$ and
$-m_Z^2 \cos 2 \beta / 3$, respectively.  The mass eigenstates
are two complex scalars ($\tilde b_1$ and $\tilde b_2$) with masses
and mixing parameter ($\sin \theta_b$) determined by diagonalizing
the matrix, Eq.~(\ref{eq:massmatrix}).  These mass eigenstates are expressed 
in terms of left-handed (L) and right-handed (R) bottom squarks, 
$\tilde{b}_L$ and $\tilde{b}_R$, as 
\label{eq:mixing}
\begin{eqnarray}
|\tilde{b}_1\rangle = \sin\theta_{b}|\tilde{b}_L\rangle + 
                                \cos\theta_{b}|\tilde{b}_R\rangle, \\
|\tilde{b}_2\rangle = \cos\theta_{b}|\tilde{b}_L\rangle - 
                                \sin\theta_{b}|\tilde{b}_R\rangle.   
\end{eqnarray}
The diagonalization of Eq.~(\ref{eq:massmatrix}) provides expressions for the 
squares of the masses of the two bottom squarks.  
%For the lighter squark ($\tilde{b}_1$) we obtain 
%\begin{equation}
%m_{{\tilde b}_1}^2
%= m_b^2 + \sin^2\theta_b (m^2_{\tilde Q} + D_L)
%        + \cos^2\theta_b (m^2_{\tilde b} + D_R)
%        + \sin 2\theta_b m_b (A_b - \mu \tan\beta).  
%\label{eq:signs}
%\end{equation}
The value of $\sin 2\theta_b$ can then be expressed in terms of the difference 
of the eigenvalues and the off-diagonal terms: 
\begin{equation}
\sin 2\theta_b
= \frac{2 m_b (A_b - \mu \tan\beta)}{m_{{\tilde b}_1}^2 - m_{{\tilde
b}_2}^2}.
\label{eq:signs}
\end{equation}
Taking $\tilde{b}_1$ to be the lighter bottom squark, we obtain the 
condition $\sin 2\theta_b (A_b - \mu \tan\beta) \le 0$.  In the limit 
in which we retain only terms enhanced by large $\tan\beta$, we determine 
that $\mu \sin 2\theta_b \ge 0$.
%The condition $\sin 2\theta_b (A_b - \mu \tan\beta) \le 0$ is necessary in 
%order that $m_{{\tilde b}_1} < m_{{\tilde b}_2}$. In the limit that we retain 
%only terms enhanced by large $\tan\beta$, we determine $\mu \sin2\theta_b \ge 0$.

There are important constraints on couplings of the bottom squarks from precise
measurements of $Z^0$ decays.  A light $\tilde b$ would be ruled out unless its
coupling to the $Z^0$ is very small.  The squark couplings to the $Z^0$ depend
on the mixing angle $\theta_b$.  As described in Ref.~\cite{Carena:2000ka}, the
lowest-order (tree-level) coupling of $\tilde{b}_1$ to the $Z^0$ can be
arranged to vanish when $\sin^2 \theta_b \sim 1/6$.  An interesting conclusion
of Ref.~\cite{Carena:2000ka} is that in order to obtain appropriately small
oblique corrections, in addition to a light bottom squark, a light top squark
with mass $\lesssim 250$ GeV is required.  In the remainder of this paper, we
use $\tilde{b}$ without a subscript to denote the lighter bottom squark.

Bottom squarks make a small contribution to the
inclusive cross section for $e^+ e^- \rightarrow$ hadrons, in comparison to
the contributions from quark production, and $\tilde{b} \tilde{b}^*$
resonances are difficult to extract from backgrounds
in $e^+ e^-$ annihilation~\cite{Nappi:1981ft}.  The angular distribution of 
hadronic jets produced in $e^+ e^-$ 
annihilation can be examined in order to bound the contribution of
scalar-quark production.  Spin-1/2 quarks and spin-0 squarks emerge with
different distributions, $(1 \pm {\rm cos}^2 \theta)$, respectively. 
Within the limits of current experimental sensitivity, the angular 
distribution measured by the CELLO
collaboration~\cite{Behrend:1986md} is consistent with the production
of a single pair of charge-1/3 squarks along with five flavors of
quark-antiquark pairs.  

The presence of a light bottom squark slows the running of the 
strong coupling strength $\alpha_s(\mu)$.
Below the gluino threshold, but above the bottom squark threshold, the 
$\beta$ function of SUSY QCD is 
\begin{equation}
\beta(\alpha_S) = \frac{\alpha_S^2}{2 \pi} \left( -11 
+ \frac{2}{3} n_f + \frac{1}{6} n_s \right) ,
\end{equation}
where $n_f$ is the number of Dirac quark flavors, and $n_s$ is the number 
of left or right squark flavors active at the scale in question.
The $\tilde{b}$ (as a color triplet scalar) contributes little to the 
running, equivalent to one quarter of a new flavor of quark and cannot be 
excluded with the current data \cite{Bethke:1994pw}. 
The exclusion by the CLEO collaboration~\cite{Savinov:2000jm} of a 
$\tilde b$ with mass 3.5 to 4.5 GeV does not apply since their analysis 
focuses only on the decays $\tilde b \rightarrow c \ell \tilde \nu$ and 
$\tilde b \rightarrow c \ell $.  The $\tilde b$ need not decay 
leptonically nor into charm.  
%The DELPHI collaboration's~\cite{Abreu:1998jy} 
%search for long-lived squarks in their $\gamma \gamma$ event sample is 
%not sensitive to $m_{\tilde b} < 15$ GeV.
%, and it does not address the 
%possibility that the bottom squarks may decay promptly via $R$-parity 
%violation into a pair of light quarks.

%%%%%%%%%%%%%%%%%%%%%%%%%%%%%%
\section{Higgs Boson Decay Rates}

In the standard model, a Higgs boson of mass below $\sim 135$ GeV, as is 
always the case for the lightest Higgs boson in the MSSM, decays predominantly
into pairs of bottom quarks, $H^0 \to b \overline{b}$.  The SM
prediction for the partial width is
\begin{equation}
\Gamma_{b} = \frac{3 g^2 m_b^2 m_h}{32 \pi m_W^2},
%\left(1 - \frac{4 m_b^2}{m_h^2}\right)^{\frac{3}{2}} 
\end{equation}
where $m_b$ is the $\overline{MS}$ bottom quark mass, 
evaluated at the mass $m_h$ of the Higgs boson, $m_W$ is the mass of the 
$W$ boson, and $g$ is the $SU(2)_W$ coupling strength.  We neglect 
the ${\cal O}(10^{-3})$ correction from the finite bottom quark mass 
in the decay phase space.  This formula is also valid for the light
SUSY Higgs boson $h$ in the decoupling limit in which the mass of the 
pseudo-scalar Higgs boson ($m_A$) is large compared to $m_Z$ and the 
couplings of $h$ with SM particles approach their SM values.  

The tree-level expression for the coupling of the lighter CP-even Higgs scalar 
$h$ to the lighter bottom squark $\tilde b$ 
is~\cite{Gunion:1984yn,Djouadi:1998az}
\bea
\label{eq:coupling}
\hspace*{-1cm}
& & \frac{g m_Z}{\cos\theta_W}
    \left[ -\frac{1}{2} \sin(\alpha+\beta)
            \left( \cos^2\theta_b - \frac{2}{3} \sin^2\theta_W \cos 2\theta_b
\right) \right. \nonumber \\
& & \qquad \qquad \left.
    + \frac{m_b^2}{m_Z^2} \frac{\sin\alpha}{\cos\beta}
    + \frac{1}{2} \sin 2\theta_b
      \frac{m_b (A_b \sin\alpha + \mu\cos\alpha)}{m_Z^2 \cos\beta}
    \right],
\eea
In this expression, $\alpha$ is the CP-even Higgs mixing angle.
In the decoupling regime, $\cos \alpha \ra \sin \beta$, 
$\sin \alpha \ra -\cos \beta$, and the term in 
Eq.~(\ref{eq:coupling}) proportional
to $\mu$ effectively grows with $\tan \beta$, unlike the coupling of $h$
to bottom quarks in the decoupling limit.  It is this feature that 
enhances the decay  $h \ra \tilde b \tilde b^*$ compared to the 
dominant SM decay, $h \ra b \overline{b}$~\cite{Carena:2000ka}.

In SUSY theories at large values of $\tan \beta$, there can be important
loop-effects (enhanced by $\tan \beta$) that affect the couplings of the
bottom quark to the Higgs bosons~\cite{Carena:1998gk}. The couplings
to the Higgs mass eigenstates may be determined from the couplings to
the weak eigenstates,
\bea
y_b \; H_d^0 \; \overline{b} b + \Delta y_b \; H_u^0 \; \overline{b} b ,
\eea
where $y_b$ is the tree-level coupling of the bottom quark to the real part
of the neutral components of the down-type Higgs 
field $H_d^0$~\cite{loopeffect}, 
and $\Delta y_b$ is the effective
coupling computed from the three-point Green function at the one-loop level. 
The one-loop effects modify the relation between the bottom quark mass and
$y_b$ by a factor potentially as large as order one,
\bea
y_b (m_b) &=& \frac{g \; m_b (m_b) }{\sqrt{2} m_W \cos \beta (1 + \Delta_b)} .
\eea
If all supersymmetric particle masses are much larger than the weak
scale, then one can neglect non-renormalizable operators in the 
effective theory below the supersymmetry breaking scale and
$\Delta_b$ is equal to  $\overline{\Delta}_b = \Delta y_b \tan \beta / y_b$. In
the presence of light sparticles, however, this relationship 
between $\Delta_b$ and $\Delta y_b$ does not hold, $\Delta_b \neq 
\overline{\Delta}_b$. (For a detailed discussion, see Ref.~\cite{Ulidavid}.)

When one moves from the weak basis to the mass basis in the
Higgs sector, and takes the so-called decoupling limit, 
in which the non-standard
heavy Higgs components become heavy, one obtains
\begin{equation}
g_{hb\overline{b}} = 
\frac{g m_b(m_h) }{ 2 m_W} \frac{ 1 + \overline{\Delta}_b}{ 1 + \Delta_b}, 
\end{equation} 
differing from the standard model coupling by a factor of order one.
Observe that in the limit of heavy supersymmetric particles, 
$\Delta_b = \overline{\Delta}_b$, and one recovers the standard model coupling.

On the other hand, for large values of $\tan\beta$, the coupling of light
bottom squarks to the light Higgs boson in the decoupling limit is governed by
the presence of a tree-level coupling of the bottom squarks to $H_u$,
\bea
g_{h \tilde b \tilde{b}^*} \simeq
\frac{g \; \mu \; m_b( m_h) \tan \beta}{2 m_W (1 + \Delta_b)} 
\sin 2 \theta_b ,
\eea
where we have neglected subdominant terms in $\tan \beta$.
Therefore, for large values of $\tan\beta$,
the width for Higgs boson decay into light bottom squarks 
may become much larger than the width for decay into bottom quarks. 
The precise relation between these decay widths depends not
only on $\tan\beta$ but also on the values of the one-loop
correction factors $\Delta_b$ and $\overline{\Delta}_b$. Because the values of
$\Delta_b$ and $\overline{\Delta}_b$
depend sensitively on the masses of other super-particles such
as the gluino, we do not attempt to include them in our results, but instead
treat them as an order-one model-dependence on our prediction for the ratio 
of the partial widths into bottom squarks and bottom quarks, 
$\Gamma_{\tilde b}/ \Gamma_b$.

\begin{figure}[ht]
\centerline{\includegraphics[height=10.0cm]{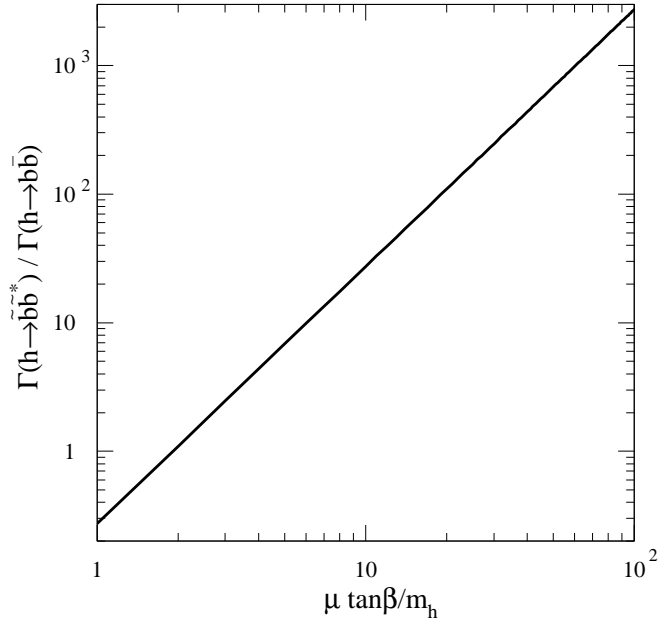}} 
\caption[]{\it The ratio of partial decay widths 
    $\Gamma(h \rightarrow \tilde{b} \tilde{b}^*) / \Gamma(h \rightarrow b \overline{b})$
    is plotted against $\mu \tan\beta/m_h$ in the limit in
    which $m_{\tilde b} \ll m_h$.}
\label{fig:plot1}
\end{figure}

The tree-level partial width for $h$ decay to a pair of $\tilde b$'s is 
\bwt
\begin{eqnarray}
\Gamma_{\tilde b} 
&=& \frac{3 g^2 m_b^2 \mu^2 \tan^2 \beta}{64 \pi m_h m_W^2} 
\sin^2 2\theta_b 
(\mu \tan \beta)^2 
\left( 1 - 4 \frac{m_{\tilde b}^2}{m_h^2} \right)^{\frac{1}{2}} , 
\end{eqnarray}
\ewt
and the ratio of partial widths is 
\begin{eqnarray}
\label{eq:ratio}
\frac{\Gamma_{\tilde b}}{\Gamma_b}
= \frac{(\mu \tan \beta)^2}{2 m_h^2} \sin^2 2\theta_b
\left( 1 - 4 \frac{m_{\tilde b}^2}{m_h^2} \right)^{\frac{1}{2}} .
%
%\approx \frac{1}{8}(\frac{\mu}{m_h})^2 \tan^2\beta \sin^2 2\theta_b ~.
\end{eqnarray}

\begin{figure}[ht]
\centerline{\includegraphics[height=10.0cm]{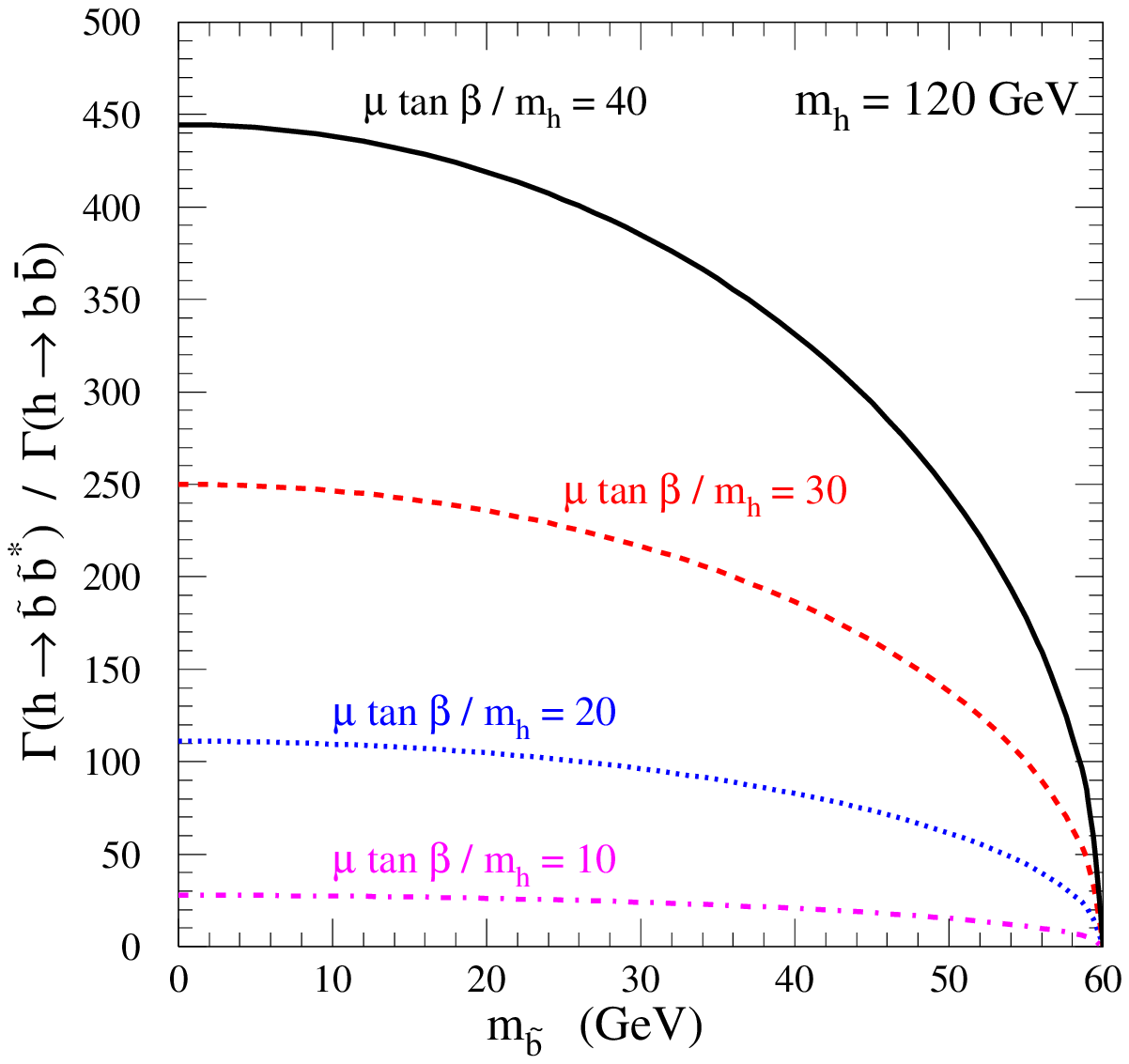}} 
\caption[]{\it The ratio of partial decay widths 
  $\Gamma(h \rightarrow \tilde{b} \tilde{b}^*) / \Gamma(h \rightarrow b \overline{b})$
    is plotted against the bottom squark mass $m_{\tilde b}$.  
    From bottom to top, the curves correspond to choices of
    $\mu \tan\beta/m_h = 10, 20, 30$ and $40$, respectively.}
\label{fig:plot2}
\end{figure}

Equation~(\ref{eq:ratio}) indicates that $(\mu \tan \beta/m_h)$ 
is the relevant quantity that determines the extent to which decays into 
bottom squarks dominate the decay process.  
The ratio as a function of $\mu \tan\beta/m_h$ is shown in 
Fig.~\ref{fig:plot1}.  We choose $m_b(m_h) = 3$ GeV, as is appropriate 
in the SM, $\sin^2 \theta_b = 1/6$, and we neglect the dependence 
on the bottom squark mass.  As stated earlier, our analysis is valid in 
the region of large $\mu \tan\beta/m_h$.  Nevertheless, we provide numerical 
results in Fig.~\ref{fig:plot1} and subsequently for values of 
$\mu \tan\beta/m_h$ that extend down to 1.  Our reason is that we take 
$\mu \tan\beta/m_h$ as a parametrization of the jet-jet rate, represented 
in our example by the $\tilde{b} \tilde{b}^*$ rate.  
We show the ratio as a function of $m_{\tilde b}$ and for a few values of 
$\mu \tan\beta/m_h$ in Fig.~\ref{fig:plot2}.  
Evident in Figs.~\ref{fig:plot1} and~\ref{fig:plot2} is that decay to a pair 
of bottom squarks is much more important than decay into bottom quarks 
%by far the dominant decay mode of the Higgs boson 
for $\mu \tan\beta/m_h > 10$, so long as $m_{\tilde b} < m_h/2$.  
%We address next the influence that bottom squarks in next-to-leading 
%order loop processes may have on other decay rates into SM particles.  
 
In addition to the direct tree level decay into bottom squark jets, 
light bottom 
squarks may affect Higgs boson decay rates into SM particles at the 
loop-level. 
Decay into two gluons occurs through loops of colored (s)particles that couple 
to the Higgs boson and to gluons.  In the SM, the only relevant contribution is 
from a loop of top quarks, with small corrections from the bottom quarks.  
In the MSSM, contributions from loops of light bottom squarks may also 
play an important role.
%Assigning the on-shell external gluons momenta $(q_1,q_2)$, Lorentz
%labels $(\mu,\nu)$, and color labels $(a,b)$, we derive the amplitude 
%\bwt
%\bea
%T^{\mu\nu}_{ab}(q_1,q_2) 
%&=& i \left( \frac{G_F}{2\sqrt{2}} \right)^{\frac12} \frac{\as}{2 \pi} 
%    m_h^2 \left( T_{SM} + T_{SUSY} \right) \nonumber \\
%&& \quad
%    \times \left( g^{\mu\nu} - \frac{2}{m_h^2} q_2^{\mu} q_1^{\nu} \right)
%    \delta_{ab} ~,
%\eea
%\ewt
%
The amplitude for $h \rightarrow g g$ is proportional to the sum 
$T_{SM} + T_{SUSY}$ where 
\bea
T_{SM}
&=& - \frac{1}{\eta_t} f(\eta_t)  ~, 
\label{eqn:tsm}  \\
T_{SUSY}
&=& - \frac{m_b \, \mu}{m_h^2} \sin 2\theta_{b} \, \tan\beta \,
    g(\eta_{\tilde b}) ~,
\label{eqn:tsusy}
\eea
and
\bea
f(x) 
&=& 1 + \frac{1-x}{x} {\rm ArcTanh}^2 \frac{\sqrt{x}}{\sqrt{x-1}} ~, \\
g(x)
&=& 1 + \frac1x {\rm ArcTanh}^2 \frac{\sqrt{x}}{\sqrt{x-1}} ~.
\eea

In Eqs.~(\ref{eqn:tsm}) and (\ref{eqn:tsusy}), 
%$\eta_t$ and $\eta_{\tilde b}$ are the ratios 
$\eta_i = {m_h^2}/{4m_i^2}$.  
%
%\be
%\eta_t = \frac{m_h^2}{4m_t^2} 
%\quad \mbox{and} \quad
%\eta_{\tilde b} = \frac{m_h^2}{4m_{\tilde b}^2} ~.
%\ee
%
Note that $\eta_t < 1$ and
$\eta_{\tilde b} > 1$, making $T_{SM}$ real but $T_{SUSY}$ complex.  
Equation~(\ref{eqn:tsusy}) shows that the sign of the SUSY 
contribution depends on the sign of the product of $\mu$ and $\sin 2\theta_b$. 
As explained after Eq.~(\ref{eq:signs}), the sign is positive.  

%The decay rate is
%
%\be
%\Gamma (h \to g g)
%= \frac{1}{32 m_h \pi} \sum_{a,b} \left| T_{ab}^{\mu\nu} \right|^2 ~. 
%\ee
%
The ratio of the total rate into the $gg$ final state, including the SUSY 
contribution, and the pure SM rate is 
\be
R = \frac{\left|T_{SM} + T_{SUSY}\right|^2}{\left|T_{SM}\right|^2}~.
\ee
\begin{figure}[t]
\centerline{\includegraphics[height=10.0cm]{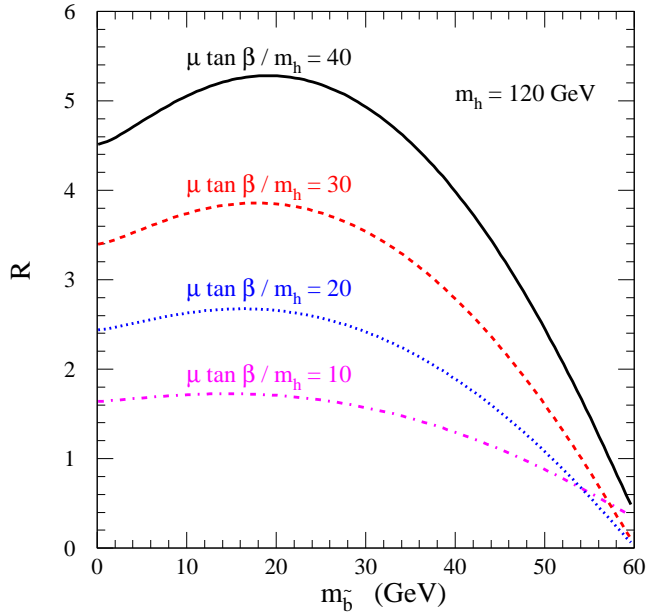}}
\caption[]{\it The ratio $R$ of the Higgs boson decay width into gluons 
divided by its SM value is plotted against $m_{\tilde
    b}$ for (from bottom left to top left) $\mu\tan\beta/m_h = 10,20,30$, 
and $40$, respectively.}
\label{fig:plot3}
\end{figure}
In Fig.~\ref{fig:plot3}, we show the dependence of $R$ on $m_{\tilde b}$ and
$\tan\beta$.  In this calculation, for completeness we include in the SM piece
the (relatively small) contribution of the bottom quark loop along with that of
the top quark loop.  We use $m_t(m_h) = 170$ GeV for the $\overline{MS}$ top
quark mass at the Higgs scale.  The relative plus sign between
Eqs.~(\ref{eqn:tsm}) and (\ref{eqn:tsusy}) leads to constructive interference
between the real parts of $T_{SM}$ and $T_{SUSY}$.  If $\mu \tan\beta/m_h =
20$, the constructive interference yields the ratio $R > 2$ for $m_{\tilde b}
\lesssim 40$ GeV.  This effect is magnified for larger values of $\mu
\tan\beta/m_h$.  The ratio $R > 1$ for a wide range of $m_{\tilde b} \lesssim
50$ GeV.  The influence of the top squark loop may be modest.  We find that the
ratio of rates for the top squark and bottom squark loops is less than 3\%, for
$5 < m_{\tilde b} < 60$ GeV (and fixed $\mu \tan \beta/m_h = 20$,
$m_{\tilde{t}} = 200$ GeV, $m_h = 120$ GeV, $A_t = 500$ GeV, and $\sin 2
\theta_{\tilde{t}} = -1$).  However, the real part of top squark contribution
is destructive with $T_{SM}$.  Inclusion of the top squark contribution reduces
the ratio $R$ in Fig.~\ref{fig:plot3} by about 10\%.  The essential content of
Fig.~\ref{fig:plot3} is that for small $m_{\tilde b}$ the bottom squark loop
can have a substantial effect on the expected rate for Higgs boson decay into a
pair of gluons.  Although the rate is enhanced, the partial width is not
magnified as much as for the $\tilde{b} \tilde{b}^*$ mode since the $g g$ decay
mode is loop-suppressed.  As shown quantitatively below, the $gg$ branching
fraction decreases as $\mu \tan \beta/m_h$ is increased, albeit less quickly
than the $b \overline{b}$ branching fraction.

For Higgs boson decay into $\gamma \gamma$, a $W$ loop contribution
dominates the SM prediction, and this remains true after SUSY contributions are
included \cite{Djouadi:1996pb}.  The contributions of the top squark, bottom
squark, and chargino are all of similar size, with specific values that depend
on the choice of parameters.  In particular, a kinematic enhancement from the
small $\tilde b$ mass is mitigated by the small charge of the $\tilde b$,
$-1/3$.  The SUSY contributions lead to a small destructive interference with
the SM loop effect in the region of bottom squark masses of interest to us,
reducing the decay rate (or the production rate in $\gamma \gamma \rightarrow
h$) by less than $10\%$ for $m_{\tilde{b}} < 30$ GeV, $20 <
\mu\tan\beta/m_h < 40$, and the same parameters for the top squark contribution
mentioned in the previous paragraph.
 
\begin{figure}[ht]
\centerline{\includegraphics[height=10.0cm]{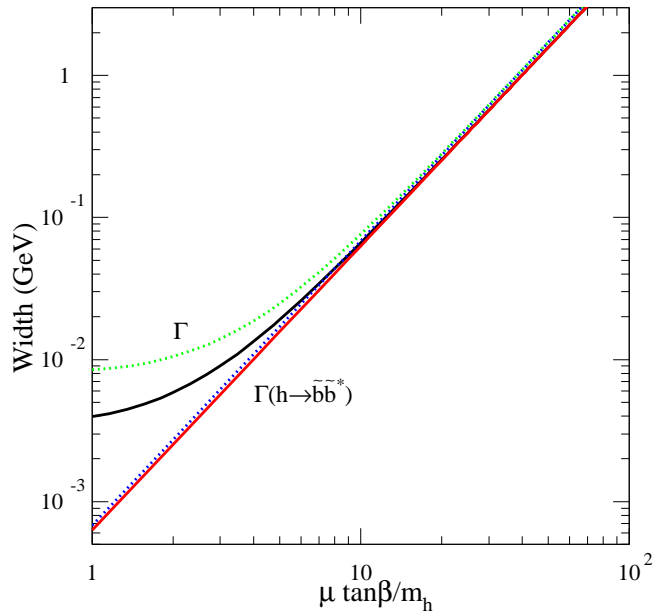}}
\caption[]{\it Total width of the Higgs boson and the partial width into 
a pair of bottom squarks as a function of the ratio $\mu \tan \beta/m_h$, 
with $m_h = 120$ GeV and $m_h = 140$ GeV.  We take $m_h \gg m_{\tilde{b}}$
For each pair of curves, the solid represents $m_h = 120$ GeV and the 
dotted $m_h = 140$ GeV.}
\label{fig:plot4}
\end{figure}

To summarize, a light bottom squark combined with moderate to large $\mu \tan
\beta/m_h$ leads the light Higgs boson of the MSSM to decay dominantly into
bottom squarks.  In contrast to the growth with $\mu \tan \beta/m_h$ of the 
partial width into bottom squarks, the partial width into bottom quarks is
affected at order one, that into gluons is enhanced with
respect to the SM, and that into
photons is relatively insensitive, being dominated by the contributions of the
$W$.  If there is a light gluino $\tilde{g}$ with mass $m_{\tilde{g}} < m_h/2$,
the channel $h \rightarrow \tilde{g} \tilde{g}$ is open~\cite{Djouadi:1994js}.
Its amplitude is provided by a loop diagram, proportional in our scenario to
the $h \tilde{b} \tilde{b}^*$ coupling and therefore enhanced by $\tan \beta$.
However, a factor of $\alpha_s$ and the usual loop suppression factors render
the final contribution small, with a branching fraction typically orders of
magnitude below that for the $\tilde{b} \tilde{b}^*$ final state.
%For small values of $\mu \tan\beta/m_h$ 
The top-squark loop contribution may also be significant for appropriate
parameters.
%competitive with the bottom-squark loop contribution considered here.  

Since the SM decay couplings are essentially unaffected, the total decay width 
of the Higgs boson is increased and the branching fractions into the SM decay 
modes are decreased accordingly.  We obtain the new total width of the Higgs boson 
by adding the SM widths (except in the $gg$ and $\gamma \gamma$ cases where we 
include the SUSY loop 
modifications) and the partial width into $\tilde{b} \tilde{b}^*$.  The total width 
and the $\tilde{b} \tilde{b}^*$ partial width are shown in 
Fig.~\ref{fig:plot4}.  For $m_h \simeq 120$ GeV, we predict a total width of 
about 250 MeV for $\mu \tan\beta/m_h = 20$, and about 1.6 GeV for 
$\mu \tan\beta/m_h = 50$.  Although the branching fraction into bottom squarks 
and the total width of the Higgs boson are enhanced in proportion to the 
square of $\mu \tan\beta/m_h$, there is no corresponding enhancement of the 
Higgs boson production rates in hadron collisions and in electron-positron 
annihilation processes, as discussed in more detail in Secs.~IV and~V.  

\begin{figure}[ht]
\centerline{\includegraphics[height=9.0cm]{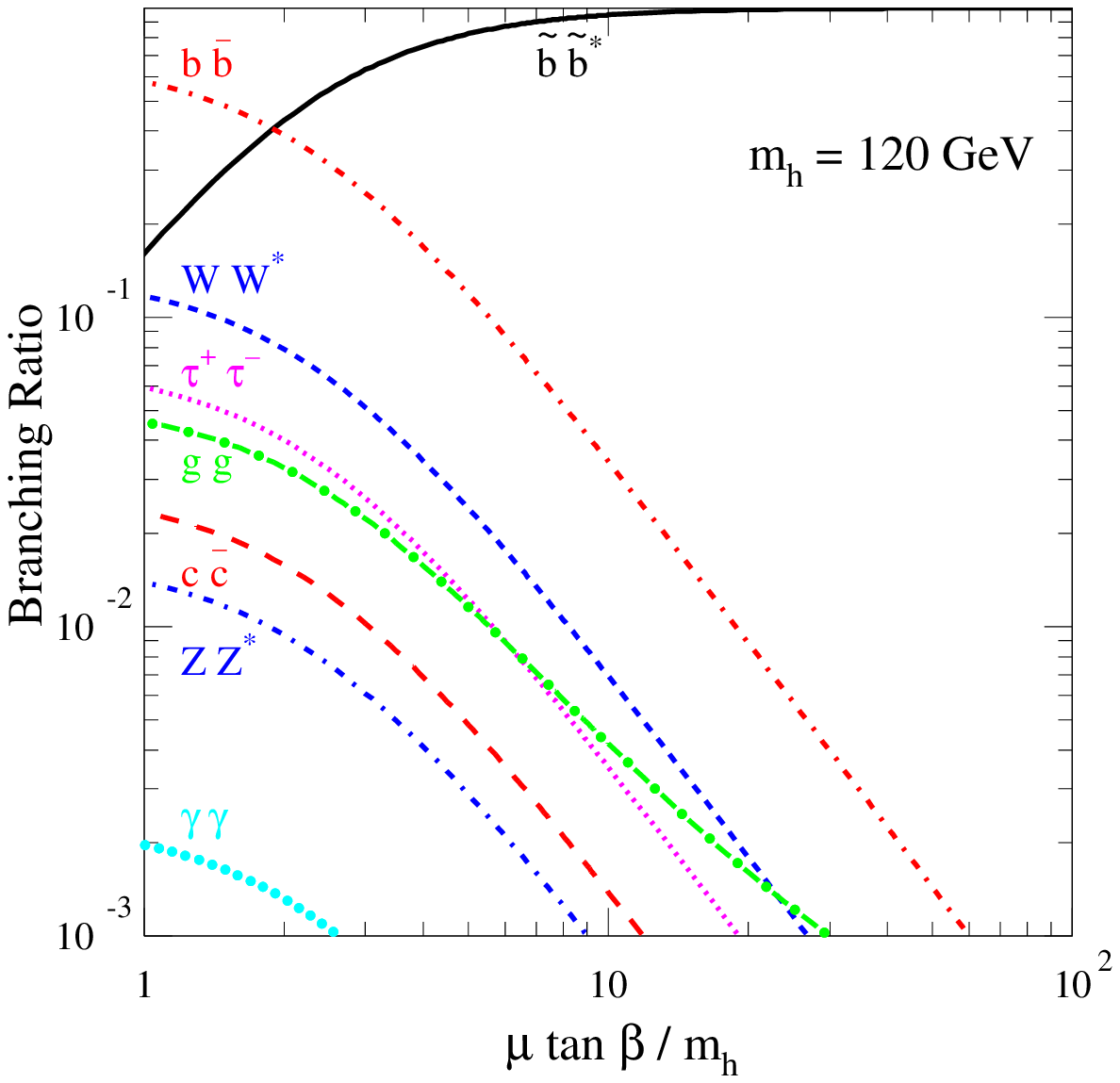}}
\centerline{\includegraphics[height=9.0cm]{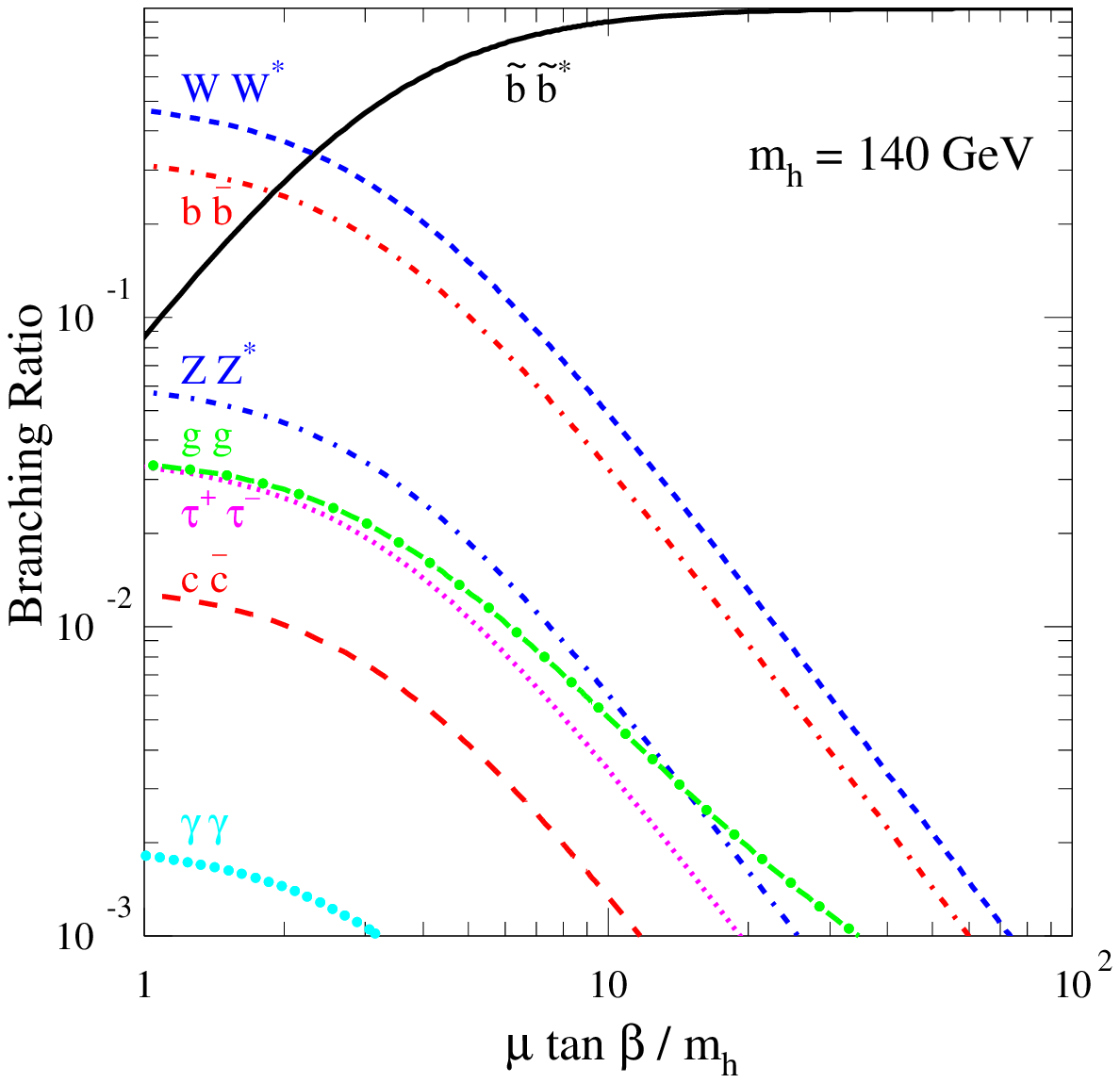}}
\caption[]{\it Branching fractions for various Higgs boson 
decay channels as a function of the ratio $\mu \tan \beta/m_h$, 
with (a) $m_h = 120$ GeV and (b) $m_h = 140$ GeV.  We fix 
$m_{\tilde{b}} = 5$ GeV in obtaining these values.}
\label{fig:plot5}
\end{figure}

A compilation of branching fractions is presented in Fig.~\ref{fig:plot5} and
in Table~I as a function of $\mu \tan\beta/m_h$.  In this table, we also
provide values of the total width, obtained after SUSY effects in the
$\tilde{b} \tilde{b}^*$, $gg$, and $\gamma \gamma$ cases are taken into
account.  We begin from the SM values of the branching
fractions~\cite{lcstudies}.  We assume that there are no SUSY corrections to
the partial widths of the $b \overline{b}$, $W W^*$, $Z Z^*$, $\tau^+ \tau^-$, and
$c \overline{c}$ modes so that the branching fractions in these cases are obtained
from the SM values of the partial widths divided by the new total width of the
Higgs boson.  In the $gg$ and $\gamma \gamma$ cases, we include the SUSY loops
effects described above in the computation of the partial widths.  At $m_h
=120$ GeV, the $b \overline{b}$ and $\tilde{b} \tilde{b}^*$ branching fractions
cross each other for $\mu \tan\beta/m_h \simeq 1.9$, where the two branching
fractions are each about 0.4.  At $m_h =140$ GeV, the $W W^*$ and $\tilde{b}
\tilde{b}^*$ branching fractions cross each other for $\mu \tan\beta/m_h \simeq
2.3$, where the two branching fractions are each about 0.34.  While $m_h =140$
GeV cannot be obtained is the usual MSSM, we provide branching ratios at this
value of mass, only slightly above that achievable in the minimal framework.

\begin{table}[ht]
\begin{tabular}{c|cccc|cccc}
\hline\hline
 & \multicolumn{8}{c}{BR $\times 10^2$} \\
\hline
$m_h$ & \multicolumn{4}{c|}{$120$ GeV} & \multicolumn{4}{c}{$140$ GeV} \\
\hline
$\mu\tan\beta / m_h$ 
  & SM & 10 & 20 & 50 & SM & 10 & 20 & 50 \\
\hline
${\tilde b} {\tilde b}^*$ 
  & 0 & $94.9$ & $98.6$ & $99.7$  
  & 0 & $90.3$ & $97.3$ & $99.5$ \\ 
$b {\overline b}$ 
  & $69$ & $3.4$ & $0.89$ & $0.14$
  & $34$ & $3.3$ & $0.88$ & $0.14$ \\ 
$W W^*$ 
  & $14$ & $0.69$ & $0.18$ & $0.029$
  & $51$ & $4.9$ & $1.3$ & $0.21$ \\ 
$Z Z^*$ 
  & $1.66$ & $0.082$ & $0.021$ & $0.003$
  & $6.3$ & $0.60$ & $0.16$ & $0.027$ \\ 
$\tau^+ \tau^-$ 
  & $7.1$ & $0.35$ & $0.091$ & $0.015$
  & $3.6$ & $0.34$ & $0.093$ & $0.015$ \\ 
$g g$ 
  & $5.2$ & $0.42$ & $0.16$ & $0.061$
  & $3.5$ & $0.51$ & $0.19$  & $0.069$ \\ 
$c {\overline c}$ 
  & $2.8$ & $0.14$ & $0.036$ & $0.006$
  & $1.4$ & $0.13$ & $0.036$ & $0.006$ \\
$\gamma \gamma$ 
  & $0.24$ & $0.011$ & $0.003$ & $0.0004$
  & $0.20$ & $0.019$ & $0.005$ & $0.0007$ \\
\hline
$\Gamma_{total}$ (MeV)
  & $3.3$ & $67$ & $257$ & $1585$
  & $7.8$ & $82$ & $303$ & $1850$ \\
\hline\hline
\end{tabular}
\vspace{6pt}
\caption{Branching ratios and total widths of the Higgs boson for masses of 
 120 and 140 GeV and $\mu\tan\beta / m_h = 10, 20, 50$.  
 We fix $m_{\tilde{b}} = 5$ GeV in obtaining these values.}
\label{tab:br}
\end{table}
%
%%%%%%%%%%%%%%%%%%%%%%%%%%%%%%
\section{Hadron Collider Phenomenology}

At the LHC, a SM-like Higgs boson of mass less than $\sim 135$ GeV 
is expected to be discovered through a variety of production processes
and decay modes \cite{atlas:1999fr},
\begin{itemize}
 \item $g g \ra h$, with $h \ra \gamma \gamma$, $h \ra W^+ W^-$, or 
       $h \ra Z Z$ ;
 \item $t \overline{t} h$, with $h \ra b \overline{b}$ or $h \ra \gamma \gamma$;
 \item $W^+ W^- (ZZ) \ra h$, with $h \ra W^+ W^-$, $h \ra \gamma \gamma$, or 
       $h \ra \tau^+ \tau^-$.
\end{itemize}
These standard searches look for Higgs boson decays into SM particles.  As 
indicated in Fig.~\ref{fig:plot5}, the presence of the light bottom squark 
suppresses the branching ratios of these decay modes by a factor of order of 
ten to several hundred, depending somewhat on $m_{\tilde b}$ and to large
degree on $\mu$ and $\tan \beta$.  This reduction raises serious questions as 
to the capability of experiments at the LHC to discover 
a Higgs boson.  The more standard decays are suppressed, and the principal 
decay mode into jets suffers from enormous QCD backgrounds.  In the analysis 
below, we assume the LHC is a $\sqrt{s} = 14$ TeV $p p$ collider with a total 
integrated luminosity of $100~{\rm fb}^{-1}$.

The gluon-gluon fusion process, $g g \ra h$ \cite{Georgi:1977gs}, is a copious
production mechanism mediated by the same loops that contribute to $h \ra g g$
described above.  In the narrow width approximation, the partonic production
cross section may be related to the decay width at leading order by the
expression 
\begin{equation}
\sigma (g g \to h)
= \frac{\pi^2}{8 m_h^3} \Gamma (h \to g g) \delta(\hat s - m_h^2)~, 
\end{equation}
and the ratio $R$ of Fig.~\ref{fig:plot3} is also
the ratio of the hadronic cross sections from this production mode in the SUSY
model to that in the SM.  As was the case for $\Gamma(h \ra g g)$, 
we conclude that for
the region of parameter space we are interested in, the $gg$ fusion production
rate is enhanced.
The large backgrounds from hadronic production of jets make discovery 
of a SM Higgs boson possible only in the distinct decay 
modes $h \ra \gamma \gamma$, $h \ra Z Z$, and $h \ra W^+ W^-$.  
In the mass range 120 to 140 GeV, the 
significance for observation at the LHC 
(signal divided by the square root of background, $S/\sqrt{B}$) 
of a SM Higgs boson is $8.1$ to $8.4$, $5.3$ to $22.1$, and 
$4.8$ to $17.7$ standard deviations 
($\sigma$'s), respectively.  A decrease in the branching fraction of any of 
these processes by a factor of 2 to 5
renders them ineffective for discovery of the $h$.  
It may be impossible to extract the new decay mode $h \ra \tilde b \tilde b^*$
from the large QCD 2-jet background, 
unless the $\tilde{b}$ has very special decay signatures.  
For $m_h \geq 150$ GeV, which cannot be realized in the MSSM but
could occur in a more general theory, decays into $W$'s and $Z$'s can
dominate over the bottom squark decays, and a high significance 
could be restored.

If there are light bottom squarks, the parton density of $\tilde{b}$
in the proton may be significant at high energies, and due to their
strong coupling to the Higgs, the partonic process 
$\tilde{b} \tilde{b}^* \ra h$ would be competitive with the
glue-glue production rate, for large enough $\tan \beta$.  This production
mode has the same experimental signature as the $gg \ra h$ mode discussed
above, and thus all of the comments regarding its observability apply 
to this process as well.  In fact, one should combine the two processes
and consider one ``inclusive'' Higgs production process.
If one assumes that the $\tilde b$ content in the proton is comparable with the
bottom content for comparable bottom quark and squark masses,
the $\tilde{b} \tilde{b}^* \ra h$ rate can be estimated from the
$b \overline b \ra h$ rate \cite{Dicus:1998hs} with an appropriate replacement
of the Higgs coupling to $b$ with the coupling to $\tilde{b}$,
and the rates at the LHC are of order 
$0.955$ to $0.576~{\rm pb} \times (\mu \tan \beta / m_h)^2$.  
This enhancement of the inclusive Higgs production, 
while growing with $(\mu \tan \beta)^2$ is compensated by the
depression of the branching fractions into observable modes, which
fall as $(\mu \tan \beta / m_h)^{-2}$.  Thus, the rate into observable
decay modes remains of the same order as in the SM.

Production in association with top quarks, $t \overline{t} h$, has a relatively
low rate because of the large masses in the final state.  In the decoupling
limit, the coupling of $h$ to top is approximately standard, so the production
rate is unaffected by the presence of a light bottom squark in the spectrum.
The expected significances for $h \ra b \overline{b}$ and $h \ra \gamma \gamma$
in the SM are approximately $9.3$ to $5.6 \sigma$ \cite{Cavalli:2002vs} and
$4.3 \sigma$ \cite{atlas:1999fr}, respectively, in the mass range of interest,
and the conclusion is that a suppression by slightly more than a factor of two
of the branching ratio $h \ra b \overline{b}$ will exclude discovery of $h$ in
this mode.  Production of $t \overline t h$ followed by the principal decay
mode $h \ra \tilde b \tilde b^*$ is expected to be difficult to observe at the
LHC because of the $t \overline t + 2$ jet background.  We estimate this
situation by considering the $h \ra b \overline{b}$ analysis of
Ref.~\cite{Cavalli:2002vs} and removing the two $b$-tag's from the Higgs decay
products as estimated in Ref.~\cite{atlas:1999fr}.  The result is that for $100
\, {\rm fb}^{-1}$ we have a significance of about $0.9 \sigma$, indicating that
$t \overline t h, h \ra $ jets is very difficult at the LHC.

The weak boson fusion modes \cite{Kauer:2000hi} can be an effective means to
search for the decays $h \ra W^+ W^-$ and $h \ra \tau^+ \tau^-$.  As in the
case of $t \overline t h$, because the couplings of $h$ to $W$ and $Z$ are
approximately standard in the decoupling limit, the production cross sections
are basically the same as in the SM, and the primary influence of the light
bottom squark is to depress the branching ratios into the distinctive decay
modes in favor of decay into jets.  The significances in the SM into the two
decay modes are $3.3$ to $13.2 \sigma$ and $10.4$ to $8.2 \sigma$ respectively
\cite{Cavalli:2002vs}, for $m_h$ in the range 120 to 140 GeV.  A reduction of
these branching fractions by a factor of 2 would prevent discovery of the Higgs
in these channels.  Weak boson fusion into a Higgs boson followed by the decay
$h \ra \tilde b \tilde b^*$ would be overwhelmed by the large QCD 4-jet
backgrounds.  For $m_h$ larger than the upper limit in the MSSM, the decay mode
into weak bosons may dominate, and discovery at the LHC would be possible with
a relatively small data sample.

\begin{figure}[ht]
\centerline{\includegraphics[height=9.0cm]{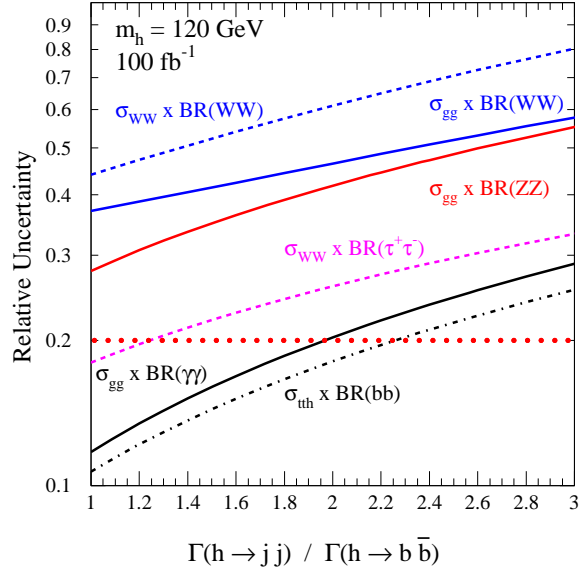}}
\centerline{\includegraphics[height=9.0cm]{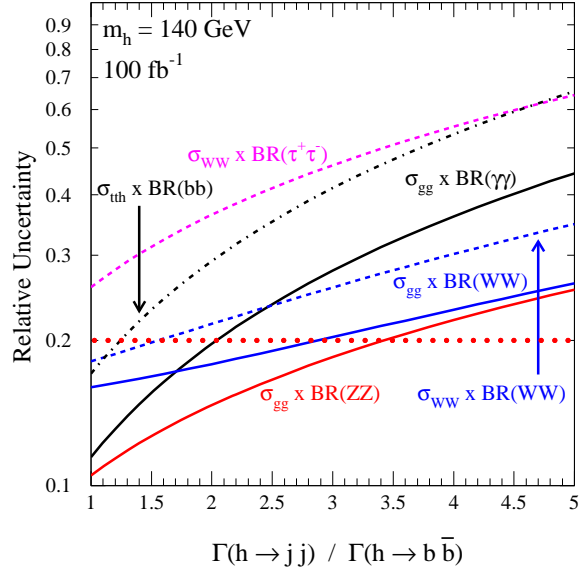}}
\caption[]{\it 
  Expected accuracy in LHC measurements of the product of production cross
  sections and branching ratios for the $WW$, $ZZ$, $b \overline{b}$, $\gamma
  \gamma$, and $\tau^+ \tau^-$ decay modes of a Higgs boson with masses 120 GeV
  and 140 GeV, as a function of the ratio of the jet-jet and the $b
  \overline{b}$ widths.  The horizontal dotted line at 0.2 indicates the
  $5\sigma$ discovery reach under the assumption $B \gg S$.  The partial widths
  for decay into $WW$, $ZZ$, $b \overline{b}$, $\gamma \gamma$, and $\tau^+
  \tau^-$ and the production cross sections are assumed to be standard. }
\label{fig:plot6}
\end{figure}

In Fig.~\ref{fig:plot6}, for $m_h = 120$ and 140 GeV, we show the accuracies
that we expect could be achieved at the LHC for measurements of the rates
(cross sections times branching ratios) of gluon fusion into a Higgs boson
followed by $h \ra \gamma \gamma$, $W^+ W^-$, and $Z Z$, and for weak boson
fusion into a Higgs boson followed by the decays $h \ra WW$ and $h \ra \tau^+
\tau^-$.  The accuracies are shown as a function of the ratio of the jet-jet
and the $b \overline {b}$ widths.  The jet-jet width is the sum of the partial
widths into $\tilde{b} \tilde{b}^*$, $b \overline{b}$, $c \overline{c}$, and
$gg$.  Note that this ratio is approximately $1.12$ in the SM, and thus the
left-most edge of the plot ($\Gamma (h \ra {\rm jets}) / \Gamma (h \ra b
\overline b) = 1$) corresponds to the case in which decay to bottom quarks is
the only hadronic decay mode of the Higgs boson.  The relative uncertainties
contain only statistical effects, $\sqrt{S + B} / S$, where we use estimates of
the backgrounds and SM signal rates presented in Refs.~\cite{Cavalli:2002vs}
and ~\cite{Zeppenfeld:2000td}.

At the LHC, it is difficult to obtain information about the Higgs boson
couplings in a model-independent way because it is impossible to observe all
possible decays in a single production mode.  One must be content with
measurements of cross sections times branching ratios and cannot make
definitive statements about the couplings themselves.  In obtaining our
estimates, we do not upgrade the glue-glue production cross section
$\sigma_{gg}$ by the SUSY bottom squark loop effects discussed in Sec.~III and
shown in Fig.~\ref{fig:plot3}.  We prefer to show the relative uncertainties as
a function of the ratio of the jet-jet and $b\overline{b}$ widths in a way that
is as model-independent as possible.
 
Experiments at the LHC can still search for the heavy SUSY Higgs bosons,
$H$, $A$, and $H^\pm$.
For example, when $\tan \beta$ is large, radiation of $A$ and $H$ from
bottom quarks is enhanced because these states couple to the bottom
quark proportionally to $\tan \beta$, even in the decoupling limit
\cite{Carena:1998gk,Dai:1994vu}.  This enhanced coupling further
insures that the branching ratio of these states into $b \overline b$
may remain competitive with the decay into $\tilde b \tilde b^*$.
The masses and properties of the heavy Higgs states are highly 
model-dependent, depending on many more of the SUSY-breaking parameters
than $m_{\tilde b}$, $\mu$, and $\tan \beta$, and thus it is difficult to
draw firm conclusions.  However, it is likely at least one of them
could be identified at the LHC with $100~{\rm fb}^{-1}$.  In this 
interesting situation, LHC experiments would discover several
elements of the Higgs sector without actually discovering the boson 
responsible for the electroweak symmetry breaking.

For Higgs boson masses between 120 and 135 GeV, searches at the Tevatron  
rely on associated production of the Higgs boson with weak bosons, 
$W$ and $Z$, and the decay mode $h \ra b \overline b$.  Discovery at the level 
of $5\sigma$ of a SM-like 
Higgs boson in this mass range is expected for integrated luminosities
greater than 20 to 60 fb$^{-1}$~\cite{Carena:2000yx}.  Depression of the Higgs 
boson branching ratio into $b \overline{b}$ will raise the required luminosities 
by the corresponding factor.  Discovery of the Higgs boson at the Tevatron 
would become very difficult under these circumstances.

Recognizing that a relatively long-lived bottom squark may pick up an 
antiquark and form a mesino, the supersymmetric partner of the $B$ 
meson, we might expect that bottom squarks will produce hadronic jets 
with leading (anti-)baryons.  R-parity violating decays of the $\tilde{b}^*$ 
may result in jets that are potentially rich in charm content.  At hadron 
colliders, the relatively small SM $c \overline{c}$ branching fraction, along with 
substantial backgrounds expected from hadronic production of gluons, followed 
by $g \rightarrow c \overline{c}$, and backgrounds from $b$ decays, 
$b \rightarrow c X$, have discouraged searches for 
$h \rightarrow c \overline{c}$.  New efforts to simulate the Higgs boson signal 
and backgrounds in the $c \overline{c}$ channel might be warranted in 
view of a possibly enhanced $c \overline{c}$ branching fraction.

\section{Phenomenology at $e^+ e^-$ Linear Colliders}
For a light Higgs boson the dominant production 
process at a lepton collider is $e^+ e^- \rightarrow Z^0 h$ via an 
intermediate $Z^0$.  Once the $Z^0$ is identified, the Higgs boson is 
discovered, independent of the Higgs boson decay modes, as a clean 
enhancement in the distribution of mass recoiling from the 
$Z^0$~\cite{lcstudies}, and the mass of the Higgs boson can be measured.
The backgrounds from the $W$ fusion and $Z^0$ fusion processes are small.  
Because the $Z h$ cross section depends on the $hZZ$ coupling strength, 
observation of the Higgs boson determines this coupling with an 
expected accuracy of $\sim 1.2$\%~\cite{lcstudies} and can establish
the Higgs boson as the principal scalar responsible for the electroweak 
symmetry breaking.  These statements, true in the SM and MSSM, remain 
valid if the Higgs boson decays primarily into a pair of bottom squarks 
since the $hZZ$ coupling is unaffected and the method does not depend on
the Higgs boson decay products.

Measurement of the Higgs boson's branching ratios is essential to establish the
properties of the boson.  For a light Higgs boson, the largest of these in the
SM is that for $h \rightarrow b \overline{b}$, with a value $\sim$ 69\% at $m_h
= 120$ GeV. At this mass, simulations~\cite{lcstudies,Brau:2001} show that the
branching fraction can determined to be $69 \pm$ 2\% based on an integrated
luminosity of 500 fb$^{-1}$ at 500 GeV.  The next largest is the branching
fraction for decay into $W W^*$, with an expected measured value of $14 \pm$
1.3\%.  As derived above, the presence of the light bottom squark decay mode
reduces the SM branching fractions into $b \overline{b}$, $W W^*$, and other
modes inversely with the square of $\mu \tan \beta/m_h$.  These values are
presented in Fig.~\ref{fig:plot5}.  In all cases except $b \overline{b}$, the
reduced branching fractions at $\mu \tan \beta/m_h = 10$ are below the
experimental accuracies estimated for a sample of data accumulated after an
integrated luminosity of 500 fb$^{-1}$ at a linear collider operating at a
center of mass energy of 500 GeV.  At $m_h = 120$ GeV, the uncertainty on the
branching fraction into the $b \overline{b}$ mode increases to 31\% when $\mu
\tan \beta/m_h =$ 10.  The branching fraction itself drops below the expected
experimental sensitivity of $\sim$ 2\% for $\mu \tan \beta/m_h >$ 13.

Determination of the $hWW$ coupling allows an experimental test of the
$SU(2)$ relationship between the $hWW$ and $hZZ$ couplings.
The usual approach for determining the $hWW$ coupling is based on measurement 
of the cross section for the $WW$ fusion process, 
$e^+ e^- \rightarrow \nu \overline{\nu} h$, plus knowledge of at least one 
branching fraction for $h$ into an observed final state. A thorough analysis 
of the expected signal and backgrounds is presented in Ref.~\cite{Desch-Meyer} 
for the $h \rightarrow b \overline{b}$ final state.  At $\sqrt{s} =$ 500 GeV and 
with an integrated luminosity of 500 fb$^{-1}$, a signal to background 
$S/B \simeq 5$ and $S/\sqrt{B} \simeq 200$ are expected.   
With the inclusion of light bottom squarks, the $b \overline{b}$ branching fraction 
drops inversely with the square of $\mu \tan \beta/m_h$, and the attendant signal 
rate falls.  Beginning from the numbers quoted in Table 3 of 
Ref.~\cite{Desch-Meyer}, and decreasing the branching fraction 
${\rm BR}(h \to b \overline{b})$, we find that 
$S/\sqrt{B}$ drops below $5$ for $\mu \tan\beta/m_h > 18$. The ratio 
$S/\sqrt{B}$ also drops to roughly $5$ in the Higgsstrahlung process 
$e^+e^- \to h Z^0 \to b \overline{b} Z^0$ for $\mu \tan\beta/m_h \approx 8$.  
Just as in the SM, the anticipated uncertainty on the determination of 
$b \overline{b}$ branching fraction dominates the overall uncertainty.  As 
$\mu \tan\beta/m_h$ is increased beyond $8$, the uncertainty in the $hWW$ 
coupling becomes greater than 10\%.  
%Since the backgrounds are significant in comparison with the signal, we 
Extrapolating from the simulation results in 
Refs.~\cite{Desch-Meyer,Brau:2001}, we expect that 
$\sigma(e^+ e^- \to h \nu \overline{\nu})$ could be determined with an accuracy 
of $\sim 36\%$ at $\sqrt{s} = 350$ GeV and $33\%$ at $\sqrt{s} = 500$ GeV 
for $\mu \tan\beta/m_h = 10$, equivalent to accuracies of $18\%$ and $17\%$, 
respectively, for the $hWW$ coupling.  

The analysis of Ref.~\cite{Desch-Meyer} can be exploited also to show 
that the Higgs boson can be discovered in the $h \rightarrow$ jet-jet 
decay channel in $e^+ e^- \rightarrow \nu \overline{\nu} h$, even at 
large $\mu \tan \beta/m_h$.  The 
dominant reducible backgrounds are listed in Table 2 of that paper.  After 
removing the ``$b$-tag'' requirement, we find that the dominant, and no-longer 
reducible backgrounds are from the processes $e^+ e^- \rightarrow e \nu W$ 
and $e^+ e^- \rightarrow e e Z$, where the $W$ and $Z$ decay to jets.  
Removing all requirements in Tables 2 and 3 of Ref.~\cite{Desch-Meyer} that 
the Higgs boson 
signal and various backgrounds proceed via the $b \overline{b}$ mode, we 
determine $S/B \simeq 0.3$ and $S/\sqrt{B} \simeq 77$ at 
$\mu \tan \beta/m_h = 10$.  The ratio $S/\sqrt{B}$ grows with 
$\mu \tan \beta/m_h$, but it saturates near 79 since the jet-jet branching 
fraction is already close to unity at $\mu \tan \beta/m_h = 10$.  

\begin{figure}[ht]
\centerline{\includegraphics[height=10.0cm]{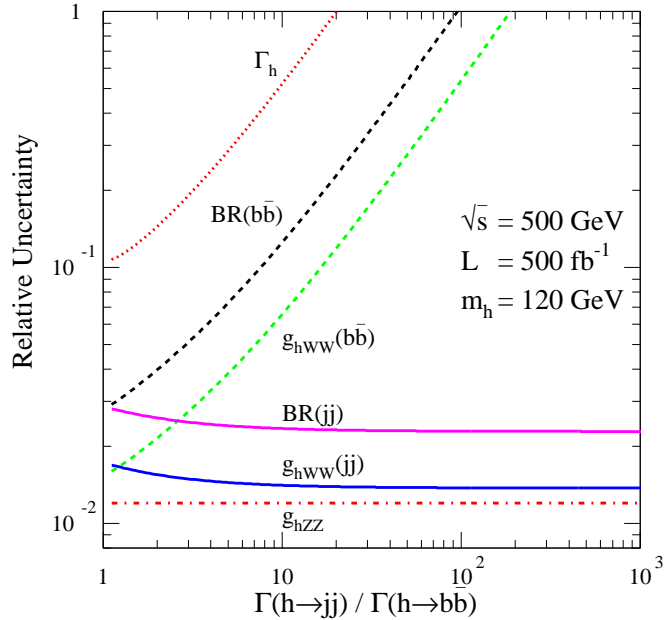}}
\caption[]{\it 
Expected accuracy in the measurements of the $b \overline{b}$ and jet-jet 
branching fractions, the $hZZ$ and $hWW$ coupling 
strengths, and the total width of the Higgs boson, as a function of the 
ratio of the jet-jet and the $b \overline {b}$ widths.  We assume the Higgs 
boson couplings to $b \overline{b}$, $ZZ$, and $WW^*$ are standard.}
\label{fig:plot7}
\end{figure}

In the weak boson fusion process, the jet-jet Higgs boson decay channel can 
also be used to determine the $hWW$ coupling at large $\mu \tan \beta/m_h$ 
with significantly greater anticipated accuracy than from the $b \overline{b}$ 
channel~\cite{explain}. In Fig.~\ref{fig:plot7}, we show the accuracies that 
we expect could be achieved in the measurements of the $b \overline{b}$ branching 
fraction, the $hZZ$ and $hWW$ coupling strengths, and the total width 
of the Higgs boson, all as a function of the ratio of the jet-jet and 
the $b \overline {b}$ widths.  We distinguish the accuracies to be expected 
for the $hWW$ coupling strength depending upon whether 
the $b \overline{b}$ or jet-jet decay mode of the Higgs boson is used.
In this plot, the jet-jet width includes the partial widths into 
$\tilde{b} \tilde{b}^*$, $b \overline{b}$, $c \overline{c}$, and $gg$.  

Knowledge of the coupling strength of the Higgs boson to the $W$, 
combined with the measurement of the Higgs boson mass,
allows one to compute the corresponding partial decay width
for $h \ra W W^*$ ($\Gamma_W$). If an independent measurement of the branching
ratio $BR(h \ra W W^*)$ is also available, one may obtain 
the Higgs boson total width ($\Gamma_h$) from the relation 
$\Gamma_h = \Gamma_W / BR(h \ra W W^*)$.  The 
accuracy on the total width is obtained from the expected accuracy on the 
determination of the branching fraction into $W W^*$ along with the    
expected accuracy on the coupling strength $g_{hWW}$.  The resulting 
uncertainty in $\Gamma_h$ is presented as a function
of the ratio of the jet-jet and the $b \overline {b}$ widths in 
Fig.~\ref{fig:plot7}.

It would be desirable to measure the total width of the Higgs 
boson {\em directly}.  For $m_h < 135$ GeV, the total width $\Gamma_h$ in 
the SM and conventional MSSM (at large $m_A$) is predicted 
to be less than 6 MeV, much too small for direct measurement at the LHC 
or at a lepton linear collider.  The substantial increase in $\Gamma_h$
arising from decays into bottom squarks may alter this expectation if 
$\mu \tan \beta/m_h$ is sufficiently large.  At $m_h = 120$ GeV and 
$\mu \tan \beta/m_h = $ 10 and 50, we expect $\Gamma_h \sim$ 66 MeV and 
1.6 GeV, respectively, both smaller than the best estimates 
of $\sim 2$ GeV for the jet-jet invariant mass resolution~\cite{lcstudies}.  
The total width will exceed 2 GeV if $\mu \tan \beta/m_h > 56$, well 
within the range of values assumed in many MSSM investigations.  
The relatively large predicted width of the Higgs boson may help 
to motivate additional effort to improve the expected jet-jet 
invariant mass resolution in order that it may be measured directly.  

The process $e^+ e^- \ra h t \overline{t}$ in which $h$ is radiated from a top
quark provides in principle the opportunity to measure the $h t \overline{t}$
coupling \cite{Juste:1999af,Baer:1999ge}.  The cross section is less than 1~fb
at $\sqrt s = 500$ GeV but increases with energy and reaches a maximum in the
$\sqrt s = 700$ to 800~GeV range.  Since $t \ra b W$ occurs with approximately
100\% branching fraction, the signal produces multi-jet final states with at
least 2 $b$ jets.  In the SM, with dominant decay of $h$ into $b \overline{b}$,
the relevant final states are $W^+ W^- b b \overline{b} \overline{b}$.  The
largest background results from gluonic radiation: $e^+ e^- \ra t \overline{t}
\ra g W^+ W^- b \overline{b}$, with $g \ra$ jet jet, and the largest
electroweak background from $e^+ e^- \ra Z^0 t \overline{t} \ra Z W^+ W^- b
\overline{b}$, with $Z \ra$ jet jet.  A 120 GeV SM Higgs boson produced at 800
GeV with an integrated luminosity of 1000 fb$^{-1}$ is considered in
Ref.~\cite{Juste:1999af}.  The signal to background is only $\sim 3$\%.  After a
neural net analysis, a potential accuracy of 5.5\% is obtained in the
determination of the $h t \overline{t}$ coupling.  A decrease of the $b
\overline{b}$ branching fraction by even a factor of 2 would seem to make
prospects untenable.  If Higgs boson decay into a pair of hadronic jets is
considered, instead of decay to $b \overline{b}$, there will be a slight
increase in the expected signal (from a branching fraction of $\sim 69$\% to
$\sim 100$\%) but the backgrounds from $g$ and $Z$ decays will increase by a
much greater factor.  It seems unlikely that the $h t \overline{t}$ coupling
could still be determined, but a full simulation of the larger signal and
backgrounds would be required for a definitive answer.

One might hope to measure the Higgs boson coupling to squarks through 
the process $e^+ e^- \ra h \tilde b \tilde b^*$, in which $h$ is radiated  
from one of the $\tilde{b}$'s, followed by the decay 
$h \rightarrow \tilde b \tilde b^*$.  However, despite the $\tan \beta$ 
enhancement of the coupling, the rate remains below $10^{-3}~{\rm fb}$ at all 
collider energies below 1 TeV, because of the small bottom squark charge
and the suppressed $P$-wave coupling to the intermediate photon.  (Recall 
that the light bottom squarks do not couple to the $Z$~\cite{Carena:2000ka}.)
These rates are small enough to preclude a single event for 
anticipated integrated luminosities, even before efficiencies are taken
into account.  For reference, after reasonable 
acceptance cuts on the jets ($p_T \geq 10$, $|y|<2$, and $\Delta R >0.4$) 
and the requirement that two of them reconstruct an invariant mass within 
5 GeV of the Higgs boson mass, the four jet background is on the order of 
$10^{-1}$ pb.

Baryon-number and R-parity violating decays of the $\tilde{b}^*$ 
into a pair of quarks, $ud$, $cd$, $us$, or $cs$~\cite{Berger:2000zk}, will 
result in jets that are potentially rich in charm content.  
Speculating that the probabilities could be equal for decay into these four 
channels, we suggest that the $c \overline{c}$ branching fraction of a 
light Higgs boson could be as great as 25\%, roughly 10 times the SM value.  
Simulations for linear collider experiments indicate that the SM
$h \rightarrow c \overline{c}$ branching fraction is expected to be determined at the 
level of $(2.8 \pm 1.1)$\%, implying that there should be no difficulty 
observing and establishing a much larger value. 

In a $\gamma \gamma$ collider~\cite{Asner:2001vh} with $\gamma$ beams produced 
by lasers backscattered from incident high energy $e^-$ beams, the coupling 
$g_{h \gamma \gamma}$ could be determined from the process 
$\gamma \gamma \rightarrow h$.  Backgrounds from light quark ($q \overline{q}$) 
production, $\gamma \gamma \ra q \overline{q} (g)$ are large, particularly if 
$q = u$ or $c$, and likely would make observation of the Higgs boson 
impossible in the jet jet decay channel.  These backgrounds can be partially 
suppressed by selections on the polarizations of the colliding photons.  
In the case of identification through the $b \overline{b}$ decay mode, the hadronic 
backgrounds include $\gamma \gamma \ra b \overline{b} (g)$ and 
$\gamma \gamma \ra c \overline{c} (g)$.  For $m_h = 120$ GeV, with the 
SM decay branching fraction into $b \overline{b}$, a simulation shows an expected 
signal of 1450 events and a background of 335 events, leading to a 
measurement of $g^2_{h \gamma \gamma} BR(h \ra b \overline{b})$ with an 
accuracy $\sqrt{S+B}/S \sim 2.9$\%~\cite{Asner:2001ia}.  This estimate 
is based on one $10^7$ seconds year of operation and excellent $b$-tagging. 
A value of $S/\sqrt{B} \sim 5$ can be maintained 
if the signal is reduced by a factor of $\sim 16$ ($S/B \simeq 0.27$). These 
numbers suggest that even for $BR(h \ra b \overline{b}) \sim 4.3$\% 
($\mu \tan \beta/m_h \sim 9$), an MSSM Higgs boson with $m_h = 120$ GeV and 
dominant decay to hadronic jets could be observed at a $\gamma \gamma$ collider, 
with an expected accuracy of $\sqrt{S+B}/S \sim 23$\% on the product 
$g^2_{h \gamma \gamma} BR(h \ra b \overline{b})$.  

%%%%%%%%%%%%%%%%%%%%%%%%%%%%%%

\section{CONCLUSIONS}

Discovery of the Higgs particle is essential to shed light on the 
mechanism of electroweak symmetry breaking.  Current strategies for 
discovery and measurement of its properties in the mass range 
$m_h < 135$ GeV rely heavily on the presumption that the principal 
branching fractions are close to those predicted in the SM or in the usual 
MSSM.  For masses in this range, the decay width of the SM Higgs boson 
is dominated by its decay into bottom quarks, $b \overline{b}$. 
In this article, we emphasize that these assumptions are unwarranted if there 
are non-standard light particles that couple weakly to the gauge 
bosons but strongly to the Higgs field. 

The small value of the bottom quark Yukawa coupling implies that the Higgs 
boson width may be modified significantly in the presence of light particles with 
relevant couplings to the Higgs boson.  In the work reported here, we analyze the 
possibility that the Higgs boson decays into new particles that manifest themselves 
as hadronic 
jets without necessarily significant bottom or charm flavor content.  As an
example of this possibility, we present the case of a light scalar bottom 
quark, with mass smaller than about 10 GeV.  While this sparticle has not been 
observed directly, its existence is consistent with all indirect experimental 
constraints.  It may decay into a pair of quarks that might be detected as 
a single energetic jet due to the high boost in Higgs boson decay.  

Under the conditions described, the decay width of the Higgs boson becomes 
several orders of magnitude larger than the width for decay into bottom 
quarks. 
For simplicity, we assume that the decay widths to standard model particles 
remain approximately constant (except in the $gg$ and $\gamma \gamma$ cases) 
and that the variation of the Higgs boson decay properties arises from the 
addition of the extra decay channel.  Decay 
into a pair of light bottom squarks dominates the total width of the Higgs 
boson for large values of the ratio of Higgs expectation values, $\tan \beta$.  
Branching fractions into standard model decay channels are reduced from their 
standard model values by a factor proportional to $\tan^{-2} \beta$.  For 
$\mu \tan \beta/m_h \simeq 13$, the $b \overline{b}$ branching fraction is reduced 
to $\sim$ 2\%.  

Experiments at the LHC are capable of looking for a Higgs boson in a variety of
channels. The Higgs boson will be found if its couplings to the $W$ and $Z$
gauge bosons and its branching ratios into bottom quarks, tau leptons, or
electroweak gauge bosons do not differ significantly from those in the SM.
These are the natural decay channels, provided there are only perturbative
modifications of the theory and no new physics below the Higgs mass scale.  We
show in this paper that for values of the branching ratio $BR(h \to j j )$
larger than two to five times that into bottom quarks, the Tevatron and the
LHC will encounter severe difficulties in finding the Higgs boson.  The
difficulty arises because the SM decay branching fractions are diminished and
the principal decay mode into a pair of hadronic jets suffers from very large
hadronic production of jet pairs.  However, experiments at these colliders are
likely to see clear evidence of low-energy supersymmetry, pointing towards the
presence of an unseen light Higgs boson in the spectrum.

Because they rely principally on the production process $e^+ e^- \rightarrow 
h Z^0$, experiments at proposed $\sqrt{s} = 500$ GeV electron-positron linear 
colliders remain fully viable for direct observation of the Higgs boson and 
measurement of its mass.  We demonstrate that this machine will discover the
Higgs particle, determine its couplings to the weak gauge bosons, and possibly 
also measure the branching ratio into bottom quarks.  The possibility of measuring 
the Higgs boson width, however, is diminished owing to the large suppression of the 
decay branching ratio into the weak gauge bosons.  If the width exceeds about 
2 GeV, a direct measurement should be possible from the invariant mass distribution 
in the jet-jet channel.  If it is smaller, determination of the width may 
have to await a Higgs boson factory based on a muon collider~\cite{muoncollider}.

In the general case considered here, the Higgs boson decays to a large extent into 
hadronic jets, possibly without definite flavor content.  Measurements of various 
properties 
of the Higgs boson, such as its full width and branching fractions, may therefore 
require a substantial improvement in the experimental jet-jet invariant mass 
resolution and a more thorough understanding of backgrounds in the 
jet-jet channel.  Full event and reconstruction studies done for the 
SM decay $h \rightarrow g g$ (where the SM branching fraction is 
$\sim$ 5\% for $m_h =120$ GeV) should be pursued further to establish 
the extent to which properties of the Higgs boson can be determined 
solely from the jet-jet mode.  

%%%%%%%%%%%%%%%%%%%%%%%%%%%%%%
\section*{ACKNOWLEDGMENTS}

The authors are grateful for conversations with James Brau, Marcela 
Carena, Debajyoti Choudhury, Stephen Mrenna, David Rainwater, 
Michael Schmitt, and Zack Sullivan.  This work was supported in part by 
the United States Department of Energy through Grant Nos.\ DE-FG02-90ER-40560 
and W-31109-ENG-38.


\begin{thebibliography}{99}

%\cite{Abbaneo:2001ix}
\bibitem{Abbaneo:2001ix}
The LEP Collaborations: ALEPH Collaboration, DELPHI Collaboration, L3 Collaboration, 
OPAL Collaboration, the LEP Electroweak Working Group, the SLD Heavy Flavour, 
Electroweak Working Group, D.~Abbaneo {\it et al.},
%``A combination of preliminary electroweak measurements and constraints 
%on the standard model,''
arXiv:hep-ex/0112021.
%%CITATION = HEP-EX 0112021;%%

\bibitem{lephiggs}
LEP Higgs Working Group for Higgs boson searches,
Proceedings of the International Europhysics Conference on High Energy
Physics (HEP 2001), Budapest, Hungary, July 12-18, 2001, 
arXiv:hep-ex/0107029 and arXiv:hep-ex/0107030.

\bibitem{Chanowitz}
M.~S.~Chanowitz,
%``The Z $\to$ anti-b b decay asymmetry: Lose-lose for the standard model,''
Phys.\ Rev.\ Lett.\  {\bf 87}, 231802 (2001)
[arXiv:hep-ph/0104024].
%%CITATION = HEP-PH 0104024;%%

%\cite{Altarelli:2001wx}
\bibitem{Altarelli:2001wx}
G.~Altarelli, F.~Caravaglios, G.~F.~Giudice, P.~Gambino, and G.~Ridolfi,
%``Indication for light sneutrinos and gauginos from precision electroweak  data,''
J.High Energy Phys. {\bf 0106}, 018 (2001)
[arXiv:hep-ph/0106029].
%%CITATION = HEP-PH 0106029;%%

\bibitem{wethree}
D.~Choudhury, T.~M.~P.~Tait, and C.~E.~M.~Wagner,
%``Beautiful mirrors and precision electroweak data,''
Phys.\ Rev.\ D {\bf 65}, 053002 (2002)
[arXiv:hep-ph/0109097].
%%CITATION = HEP-PH 0109097;%

\bibitem{Peskin}
M.~E.~Peskin and J.~D.~Wells,
%``How can a heavy Higgs boson be consistent with the precision  electroweak measurements?,''
Phys.\ Rev.\ D {\bf 64}, 093003 (2001)
[arXiv:hep-ph/0101342].
%%CITATION = HEP-PH 0101342;%%

%\cite{Chivukula:2000px}
\bibitem{Chivukula:2000px}
H.~Georgi,
%``Effective field theory and electroweak radiative corrections,''
Nucl.\ Phys.\ B {\bf 363}, 301 (1991);
%%CITATION = NUPHA,B363,301;%%
B.~A.~Dobrescu and C.~T.~Hill,
%``Electroweak symmetry breaking via top condensation seesaw,''
Phys.\ Rev.\ Lett.\  {\bf 81}, 2634 (1998)
[arXiv:hep-ph/9712319];
%%CITATION = HEP-PH 9712319;%%
R.~S.~Chivukula, B.~A.~Dobrescu, H.~Georgi, and C.~T.~Hill,
%``Top quark seesaw theory of electroweak symmetry breaking,''
Phys.\ Rev.\ D {\bf 59}, 075003 (1999)
[arXiv:hep-ph/9809470];
%%CITATION = HEP-PH 9809470;%%
H.~J.~He, T.~Tait, and C.~P.~Yuan,
%``New topflavor models with seesaw mechanism,''
Phys.\ Rev.\ D {\bf 62}, 011702 (2000)
[arXiv:hep-ph/9911266];
%%CITATION = HEP-PH 9911266;%%
H.~Collins, A.~K.~Grant, and H.~Georgi,
%``The phenomenology of a top quark seesaw model,''
Phys.\ Rev.\ D {\bf 61}, 055002 (2000)
[arXiv:hep-ph/9908330];
%%CITATION = HEP-PH 9908330;%%
H.~J.~He, N.~Polonsky, and S.~F.~Su,
%``Extra families, Higgs spectrum and oblique corrections,''
Phys.\ Rev.\ D {\bf 64}, 053004 (2001)
[arXiv:hep-ph/0102144];
%%CITATION = HEP-PH 0102144;%%
R.~S.~Chivukula, C.~Hoelbling, and N.~Evans,
%``Limits on a composite Higgs boson,''
Phys.\ Rev.\ Lett.\  {\bf 85}, 511 (2000)
[arXiv:hep-ph/0002022]; 
%%CITATION = HEP-PH 0002022;%%
H.~J.~He, C.~T.~Hill, and T.~M.~P.~Tait,
%``Top quark seesaw, vacuum structure and electroweak precision  constraints,''
Phys.\ Rev.\ D {\bf 65}, 055006 (2002)
[arXiv:hep-ph/0108041]. 
%%CITATION = HEP-PH 0108041;%%

\bibitem{wethreeII}
D.~Choudhury, T.~M.~P.~Tait, and C.~E.~M.~Wagner,
%``Probing heavy Higgs boson models with a TeV linear collider,''
Phys. \ Rev. \ D, in press, arXiv:hep-ph/0202162.
%%CITATION = HEP-PH 0202162;%%

%\cite{Gunion:1984yn}
\bibitem{Gunion:1984yn}
J.~F.~Gunion and H.~E.~Haber,
%``Higgs Bosons In Supersymmetric Models. 1,''
Nucl.\ Phys.\ B {\bf 272}, 1 (1986)
[Erratum-ibid.\ B {\bf 402}, 567 (1986)].
%%CITATION = NUPHA,B272,1;%%

%\cite{Heinemeyer:1998jw}
\bibitem{Heinemeyer:1998jw}
S.~Heinemeyer, W.~Hollik, and G.~Weiglein,
%``{QCD} corrections to the masses of the neutral CP-even Higgs bosons in  the MSSM,''
Phys.\ Rev.\ D {\bf 58}, 091701 (1998)
[arXiv:hep-ph/9803277];
%%CITATION = HEP-PH 9803277;%%
%\cite{Heinemeyer:1998kz}
%\bibitem{Heinemeyer:1998kz}
S.~Heinemeyer, W.~Hollik, and G.~Weiglein,
%``Precise prediction for the mass of the lightest Higgs boson in the  MSSM,''
Phys.\ Lett.\ B {\bf 440}, 296 (1998)
[arXiv:hep-ph/9807423];
%%CITATION = HEP-PH 9807423;%%
%\cite{Heinemeyer:1998np}
%\bibitem{Heinemeyer:1998np}
S.~Heinemeyer, W.~Hollik, and G.~Weiglein,
%``The masses of the neutral CP-even Higgs bosons in the MSSM: Accurate  analysis at the two-loop level,''
Eur.\ Phys.\ J.\ C {\bf 9}, 343 (1999)
[arXiv:hep-ph/9812472];
%%CITATION = HEP-PH 9812472;%%
%\cite{Carena:1995wu}
%\bibitem{Carena:1995wu}
M.~Carena, M.~Quiros, and C.~E.~M.~Wagner,
%``Effective potential methods and the Higgs mass spectrum in the MSSM,''
Nucl.\ Phys.\ B {\bf 461}, 407 (1996)
[arXiv:hep-ph/9508343];
%%CITATION = HEP-PH 9508343;%%
H.~E.~Haber, R.~Hempfling, and A.~H.~Hoang,
%``Approximating the radiatively corrected Higgs mass in the minimal
% supersymmetric model,''
Z.\ Phys.\ C {\bf 75}, 539 (1997)
[arXiv:hep-ph/9609331]; 
%%CITATION = HEP-PH 9609331;%%
J.~R.~Espinosa and R.~J.~Zhang,
%``MSSM lightest CP-even Higgs boson mass to O(alpha(s) alpha(t)):  The
% effective potential approach,''
J. High Energy Phys. {\bf 0003}, 026 (2000)
[arXiv:hep-ph/9912236];
%%CITATION = HEP-PH 9912236;%%
J.~R.~Espinosa and R.~J.~Zhang,
%``Complete two-loop dominant corrections to the mass of the lightest
% CP-even Higgs boson in the minimal supersymmetric standard model,''
Nucl.\ Phys.\ B {\bf 586}, 3 (2000)
[arXiv:hep-ph/0003246];
%%CITATION = HEP-PH 0003246;%%
M.~Carena, H.~E.~Haber, S.~Heinemeyer, W.~Hollik, C.~E.~M.~Wagner, and
G.~Weiglein,
%``Reconciling the two-loop diagrammatic and effective field theory
% computations of the mass of the lightest CP-even Higgs boson in the
% MSSM,''
Nucl.\ Phys.\ B {\bf 580}, 29 (2000)
[arXiv:hep-ph/0001002];
%%CITATION = HEP-PH 0001002;%%
G.~Degrassi, P.~Slavich, and F.~Zwirner,
%``On the neutral Higgs boson masses in the MSSM for arbitrary stop
% mixing,''
Nucl.\ Phys.\ B {\bf 611}, 403 (2001)
[arXiv:hep-ph/0105096];
%%CITATION = HEP-PH 0105096;%%
A.~Brignole, G.~Degrassi, P.~Slavich, and F.~Zwirner,
%``On the O(alpha**2(t)) two-loop corrections to the neutral Higgs boson
% masses in the MSSM,''
arXiv:hep-ph/0112177.
%%CITATION = HEP-PH 0112177;%%

\bibitem{hdecay} We use the HDECAY code to compute the total width in 
the SM; 
A.~Djouadi, J.~Kalinowski, and M.~Spira, 
%``HDECAY: a Program for Higgs Boson Decays in the 
%Standard Model and its Supersymmetric Extension''
Comput.Phys.Commun. {\bf 108}, 56 (1998) [arXiv: hep-ph/9704448].   

\bibitem{martinwells}
S.~P.~Martin and J.~D.~Wells,
%``Motivation and detectability of an invisibly-decaying Higgs boson at the Fermilab Tevatron,''
Phys.\ Rev.\ D {\bf 60}, 035006 (1999)
[arXiv:hep-ph/9903259];
%%CITATION = HEP-PH 9903259;%%
J.~D.~Wells, arXiv:hep-ph/0205328.
%%CITATION = HEP-PH 0205328;%%

\bibitem{flavorind}
LEP Higgs Working Group for Higgs boson searches,
%``Flavor independent search for hadronically decaying neutral Higgs  bosons at LEP,''
arXiv:hep-ex/0107034.
%%CITATION = HEP-EX 0107034;%%

%\cite{Carena:2000ka}
\bibitem{Carena:2000ka}
M.~Carena, S.~Heinemeyer, C.~E.~M.~Wagner, and G.~Weiglein,
%``Do electroweak precision data and Higgs mass constraints rule out a  scalar bottom quark with mass of O(5-GeV)?,''
Phys.\ Rev.\ Lett.\  {\bf 86}, 4463 (2001)
[arXiv:hep-ph/0008023].
%%CITATION = HEP-PH 0008023;%%

%\cite{Berger:2000mp}
\bibitem{Berger:2000mp}
E.~L.~Berger, B.~W.~Harris, D.~E.~Kaplan, Z.~Sullivan, T.~M.~P.~Tait, 
and C.~E.~M.~Wagner,
%``Low energy supersymmetry and the Tevatron bottom-quark cross section,''
Phys.\ Rev.\ Lett.\  {\bf 86}, 4231 (2001)
[arXiv:hep-ph/0012001].
%%CITATION = HEP-PH 0012001;%%

%\cite{Berger:2001jb}
\bibitem{Berger:2001jb}
E.~L.~Berger and L.~Clavelli,
%``Upsilon decay to a pair of bottom squarks,''
Phys.\ Lett.\ B {\bf 512}, 115 (2001)
[arXiv:hep-ph/0105147]; 
%%CITATION = HEP-PH 0105147;%%
%\cite{Berger:2002gu}
%\bibitem{Berger:2002gu}
E.~L.~Berger and J.~Lee,
%``Hadronic decays of chi/bJ into light bottom squarks,''
arXiv:hep-ph/0203092, Phys.~Rev.~D, in press.  
%%CITATION = HEP-PH 0203092;%%

%\cite{Berger:2002kc}
\bibitem{Berger:2002kc}
E.~L.~Berger,
%``The puzzle of the bottom quark production cross section,''
arXiv:hep-ph/0201229, Int.~J.~Mod.~Phys.~A, in press.
%%CITATION = HEP-PH 0201229;%%

%\cite{Dedes:2000nv}
\bibitem{Dedes:2000nv}
A.~Dedes and H.~K.~Dreiner,
%``A light bottom squark in the MSSM,''
J. High Energy Phys. {\bf 0106}, 006 (2001)
[arXiv:hep-ph/0009001].  
%%CITATION = HEP-PH 0009001;%%

%\cite{Nierste:2000ez}
\bibitem{Nierste:2000ez}
U.~Nierste and T.~Plehn,
%``Probing light sbottoms with B decays,''
Phys.\ Lett.\ B {\bf 493}, 104 (2000)
[arXiv:hep-ph/0008321].
%%CITATION = HEP-PH 0008321;%%

%\cite{Becher:2001zb}
\bibitem{Becher:2001zb}
T.~Becher, S.~Braig, M.~Neubert, and A.~Kagan,
%``Flavor-change with ultra-light sbottom and gluinos,''
arXiv:hep-ph/0112129;
%%CITATION = HEP-PH 0112129;
%\cite{Becher:2002ue}
%\bibitem{Becher:2002ue}
T.~Becher, S.~Braig, M.~Neubert and A.~L.~Kagan,
%``Constraints on Light Bottom Squarks from Radiative B-Meson Decays,''
arXiv:hep-ph/0205274.
%%CITATION = HEP-PH 0205274;%%

%\cite{Cao:2001rz}
\bibitem{Cao:2001rz}
J.~J.~Cao, Z.~H.~Xiong, and J.~M.~Yang,
%``Can MSSM with light sbottom and light gluino survive Z-peak  constraints?,''
Phys.\ Rev.\ Lett.\  {\bf 88}, 111802 (2002)
[arXiv:hep-ph/0111144]; 
%%CITATION = HEP-PH 0111144;%%
%\cite{Cho:2002mt}
%\bibitem{Cho:2002mt}
G.~C.~Cho,
%``Light bottom squark and gluino confront electroweak precision  measurements,''
arXiv:hep-ph/0204348; 
%%CITATION = HEP-PH 0204348;%%
%\cite{Baek:2002xf}
%\bibitem{Baek:2002xf}
S.~Baek,
%``Very light sbottom and gluino scenario confronting electroweak precision tests,''
arXiv:hep-ph/0205013.
%%CITATION = HEP-PH 0205013;%%

%\cite{Leibovich:2002qp}
\bibitem{Leibovich:2002qp}
A.~K.~Leibovich and D.~Rainwater,
%``Increased yield of t anti-t b anti-b at hadron colliders in low-energy  supersymmetry,''
arXiv:hep-ph/0202174.
%%CITATION = HEP-PH 0202174;%%

%\cite{Nappi:1981ft}
\bibitem{Nappi:1981ft}
C.~R.~Nappi,
%``Spin 0 Quarks In E+ E- Annihilation,''
Phys.\ Rev.\ D {\bf 25}, 84 (1982);
%%CITATION = PHRVA,D25,84;%%
%\cite{Pacetti:2000re}
%\bibitem{Pacetti:2000re}
S.~Pacetti and Y.~Srivastava,
%``Resolution of a long standing discrepancy in R with spin zero quarks,''
arXiv:hep-ph/0007318.
%%CITATION = HEP-PH 0007318;%%

%\cite{Behrend:1986md}
\bibitem{Behrend:1986md}
H.~J.~Behrend {\it et al.}  [CELLO Collaboration],
%``Determination Of Alpha-S And Sin**2theta(W) From Measurements Of The Total Hadronic Cross-Section In E+ E- Annihilation,''
Phys.\ Lett.\ B {\bf 183}, 400 (1987).
%%CITATION = PHLTA,B183,400;%%

%\cite{Savinov:2000jm}
\bibitem{Savinov:2000jm}
V.~Savinov {\it et al.}  [CLEO Collaboration],
%``Search for a scalar bottom quark with mass 3.5-GeV/c**2 to  4.5-GeV/c**2,''
Phys.\ Rev.\ D {\bf 63}, 051101 (2001)
[arXiv:hep-ex/0010047].
%%CITATION = HEP-EX 0010047;%%

%\cite{Abreu:1998jy}
%\bibitem{Abreu:1998jy}
%P.~Abreu {\it et al.}  [DELPHI Collaboration],
%``A search for heavy stable and longlived squarks and sleptons in e+ e-  collisions at energies from 130-GeV to 183-GeV,''
%Phys.\ Lett.\ B {\bf 444}, 491 (1998)
%[arXiv:hep-ex/9811007].
%%CITATION = HEP-EX 9811007;%%

%\cite{Berger:2000zk}
\bibitem{Berger:2000zk}
E.~L.~Berger, B.~W.~Harris, and Z.~Sullivan,
%``Direct probes of R-parity-violating supersymmetric couplings via  single-top-squark production,''
Phys.\ Rev.\ D {\bf 63}, 115001 (2001)
[arXiv:hep-ph/0012184], 
%%CITATION = HEP-PH 0012184;%%
%\cite{Berger:1999zt}
%\bibitem{Berger:1999zt}
%E.~L.~Berger, B.~W.~Harris and Z.~Sullivan,
%``Single-top-squark production via R-parity-violating supersymmetric  couplings in hadron collisions,''
Phys.\ Rev.\ Lett.\  {\bf 83}, 4472 (1999)
[arXiv:hep-ph/9903549], and references therein.  
%%CITATION = HEP-PH 9903549;%% 

\bibitem{cosmological}
We assume that the lifetime of the bottom squark is less than the cosmological 
time scale so that these squarks make no contribution to the dark matter density.  

%\cite{Bethke:1994pw}
\bibitem{Bethke:1994pw}
S.~Bethke,
%``Summary of alpha(s) measurements,''
Nucl.\ Phys.\ Proc.\ Suppl.\  {\bf 39BC}, 198 (1995).
%%CITATION = NUPHZ,39BC,198;%%

%\cite{Djouadi:1998az}
\bibitem{Djouadi:1998az}
A.~Djouadi,
%``Squark effects on Higgs boson production and decay at the LHC,''
Phys.\ Lett.\ B {\bf 435}, 101 (1998)
[arXiv:hep-ph/9806315].
%%CITATION = HEP-PH 9806315;%%

%\cite{Carena:1998gk}
\bibitem{Carena:1998gk}
M.~Carena, S.~Mrenna, and C.~E.~M.~Wagner,
%``MSSM Higgs boson phenomenology at the Tevatron collider,''
Phys.\ Rev.\ D {\bf 60}, 075010 (1999)
[arXiv:hep-ph/9808312].
%%CITATION = HEP-PH 9808312;%%

\bibitem{loopeffect}
We ignore a small one-loop correction that affects the coupling of 
bottom quarks to $H_d$.

\bibitem{Ulidavid}
M.~Carena, D.~Garcia, U.~Nierste, and C.~E.~M.~Wagner,
%``Effective Lagrangian for the anti-t b H+ interaction in the MSSM and  
%charged Higgs phenomenology,''
Nucl.\ Phys.\ B {\bf 577}, 88 (2000)
[arXiv:hep-ph/9912516].
%%CITATION = HEP-PH 9912516;%%

%\cite{Djouadi:1996pb}
\bibitem{Djouadi:1996pb}
A.~Djouadi, V.~Driesen, W.~Hollik, and J.~I.~Illana,
%``The coupling of the lightest SUSY Higgs boson to two photons in the  decoupling regime,''
Eur.\ Phys.\ J.\ C {\bf 1}, 149 (1998)
[arXiv:hep-ph/9612362].
%%CITATION = HEP-PH 9612362;%%

%\cite{Djouadi:1994js}
\bibitem{Djouadi:1994js}
A.~Djouadi and M.~Drees,
%``Higgs and Z boson decays into light gluinos,''
Phys.\ Rev.\ D {\bf 51}, 4997 (1995)
[arXiv:hep-ph/9411314].
%%CITATION = HEP-PH 9411314;%%

\bibitem{lcstudies}
American Linear Collider Working Group, T. Abe {\em et al}, ``Linear 
Collider Physics Resource Book for Snowmass 2001", SLAC-R-570; TESLA 
Technical Design Report, Ed. by R.~Heuer, D.~Miller, F.~Richard, 
A.~Wagner, and P.~Zerwas, www.desy.de/~lcnotes/tdr.

%\cite{atlas:1999fr}
\bibitem{atlas:1999fr}
ATLAS Collaboration,
%``ATLAS detector and physics performance. 
Technical design report, Vol. 2, 
CERN/LHCC/99-15 (1999); 
CMS Collaboration, Technical Proposal, report CERN/LHCC/94-38 (1994).

%\cite{Georgi:1977gs}
\bibitem{Georgi:1977gs}
H.~M.~Georgi, S.~L.~Glashow, M.~E.~Machacek, and D.~V.~Nanopoulos,
%``Higgs Bosons From Two Gluon Annihilation In Proton-Proton Collisions,''
Phys.\ Rev.\ Lett.\  {\bf 40}, 692 (1978).
%%CITATION = PRLTA,40,692;%%

%\cite{Dicus:1998hs}
\bibitem{Dicus:1998hs}
D.~Dicus, T.~Stelzer, Z.~Sullivan and S.~Willenbrock,
%``Higgs boson production in association with bottom quarks at  next-to-leading order,''
Phys.\ Rev.\ D {\bf 59}, 094016 (1999)
[arXiv:hep-ph/9811492];
%%CITATION = HEP-PH 9811492;%%
%\cite{Campbell:2002zm}
%\bibitem{Campbell:2002zm}
J.~Campbell, R.~K.~Ellis, F.~Maltoni and S.~Willenbrock,
%``Higgs-boson production in association with a single bottom quark,''
arXiv:hep-ph/0204093.
%%CITATION = HEP-PH 0204093;%%

%\cite{Cavalli:2002vs}
\bibitem{Cavalli:2002vs}
D.~Cavalli {\it et al.},
``The Higgs working group: Summary report'', 
To appear in the proceedings of Workshop on Physics at TeV Colliders, 
Les Houches, France, 21 May - 1 Jun 2001,  
arXiv:hep-ph/0203056.
%%CITATION = HEP-PH 0203056;%%

%\cite{Kauer:2000hi}
\bibitem{Kauer:2000hi}
N.~Kauer, T.~Plehn, D.~Rainwater, and D.~Zeppenfeld,
%``H $\to$ W W as the discovery mode for a light Higgs boson,''
Phys.\ Lett.\ B {\bf 503}, 113 (2001)
[arXiv:hep-ph/0012351];
%%CITATION = HEP-PH 0012351;%%
%\cite{Plehn:1999xi}
%\bibitem{Plehn:1999xi}
T.~Plehn, D.~Rainwater, and D.~Zeppenfeld,
%``A method for identifying H $\to$ tau tau $\to$ e+- mu-+ missing p(T)  at the CERN LHC,''
Phys.\ Rev.\ D {\bf 61}, 093005 (2000)
[arXiv:hep-ph/9911385];
%%CITATION = HEP-PH 9911385;%%
%\cite{Rainwater:1997dg
%\bibitem{Rainwater:1997dg}
D.~Rainwater and D.~Zeppenfeld,
%``Searching for H $\to$ gamma gamma in weak boson fusion at the LHC,''
JHEP {\bf 9712}, 005 (1997)
[arXiv:hep-ph/9712271].
%%CITATION = HEP-PH 9712271;%%

%\cite{Zeppenfeld:2000td}
\bibitem{Zeppenfeld:2000td}
D.~Zeppenfeld, R.~Kinnunen, A.~Nikitenko, and E.~Richter-Was,
%``Measuring Higgs boson couplings at the LHC,''
Phys.\ Rev.\ D {\bf 62}, 013009 (2000)
[arXiv:hep-ph/0002036].
%%CITATION = HEP-PH 0002036;%%

%\cite{Dai:1994vu}
\bibitem{Dai:1994vu}
J.~Dai, J.~F.~Gunion, and R.~Vega,
%``LHC detection of neutral MSSM Higgs bosons via g g $\to$ b anti-b h $\to$ b anti-b b anti-b,''
Phys.\ Lett.\ B {\bf 345}, 29 (1995)
[arXiv:hep-ph/9403362];
%%CITATION = HEP-PH 9403362;%%
%\cite{Dai:1996rn}
%\bibitem{Dai:1996rn}
J.~Dai, J.~F.~Gunion, and R.~Vega,
%``Detection of neutral MSSM Higgs bosons in four-b final states at the  Tevatron and the LHC: An update,''
Phys.\ Lett.\ B {\bf 387}, 801 (1996)
[arXiv:hep-ph/9607379];
%%CITATION = HEP-PH 9607379;%%
%\cite{Diaz-Cruz:1998qc}
%\bibitem{Diaz-Cruz:1998qc}
J.~L.~Diaz-Cruz, H.~J.~He, T.~M.~P.~Tait, and C.~P.~Yuan,
%``Higgs bosons with large bottom Yukawa coupling at Tevatron and LHC,''
Phys.\ Rev.\ Lett.\  {\bf 80}, 4641 (1998)
[arXiv:hep-ph/9802294];
%%CITATION = HEP-PH 9802294;%%
%\cite{Balazs:1998nt}
%\bibitem{Balazs:1998nt}
C.~Balazs, J.~L.~Diaz-Cruz, H.~J.~He, T.~M.~P.~Tait, and C.~P.~Yuan,
%``Probing Higgs bosons with large bottom Yukawa coupling at hadron  colliders,''
Phys.\ Rev.\ D {\bf 59}, 055016 (1999)
[arXiv:hep-ph/9807349];
%%CITATION = HEP-PH 9807349;%%
%\cite{Campbell:2002zm}
%\bibitem{Campbell:2002zm}
J.~Campbell, R.~K.~Ellis, F.~Maltoni, and S.~Willenbrock,
%``Higgs-boson production in association with a single bottom quark,''
arXiv:hep-ph/0204093.
%%CITATION = HEP-PH 0204093;%%

%\cite{Carena:2000yx}
\bibitem{Carena:2000yx}
M.~Carena {\it et al.},
%``Report of the Tevatron Higgs working group,''
arXiv:hep-ph/0010338.
%%CITATION = HEP-PH 0010338;%%

%\cite{Brau:2000bj}
\bibitem{Brau:2001}
J.~E.~Brau, C.~Potter, and M.~Iwasaki, \\
$\rm{http://www.slac.stanford.edu/econf/C010630/forweb/P118\_potter.pdf}$ 
to be found among the papers prepared for the proceedings of the 2001 
Snowmass Summer Study, 
$\rm{http://www.slac.stanford.edu/econf/C010630/forweb/paperstats.html}$.
%Batavia 2000, Physics and
%experiments with future linear $e^+ e^-$ colliders, 795.
%``Higgs branching ratio measurements at a future linear collider,''.

\bibitem{Desch-Meyer}
K.~Desch and N.~Meyer, LC-PHSM-2001-25.

\bibitem{explain}
To obtain the uncertainty on the branching fraction into a pair of jets, 
we must first estimate the number of signal and background 
events in the jet-jet channel.  We begin with the numbers presented in 
Ref.~\cite{Brau:2001} of simulated signal and background events 
for $h \rightarrow b \overline{b}$ in the process $e^+ e^- \rightarrow Z h$.  We 
remove the $b$-tag requirement, increasing both signal and background by 
$(1/0.75)^2$.  The signal sample is then multiplied by (1/0.69), the inverse 
of the SM $b \overline{b}$ branching fraction, and by the jet-jet  
branching fraction in our model.  Since the primary background arises from 
$Z$ decay, the background is increased by $1/R_b \simeq 1/0.21$, where  
$R_b$ is the measured fraction of the hadronic width of the $Z$ into 
$b\overline{b}$.   

%\cite{Juste:1999af}
\bibitem{Juste:1999af}
A.~Juste and G.~Merino,
%``Top-Higgs Yukawa coupling measurement at a linear e+ e- collider,''
arXiv:hep-ph/9910301.
%%CITATION = HEP-PH 9910301;%%
  
%\cite{Baer:1999ge}
\bibitem{Baer:1999ge}
H.~Baer, S.~Dawson and L.~Reina,
%``Measuring the top quark Yukawa coupling at a linear e+ e- collider,''
Phys.\ Rev.\ D {\bf 61}, 013002 (2000)
[arXiv:hep-ph/9906419];
%%CITATION = HEP-PH 9906419;%%
%\cite{Dittmaier:1998dz}
%\bibitem{Dittmaier:1998dz}
S.~Dittmaier, M.~Kramer, Y.~Liao, M.~Spira and P.~M.~Zerwas,
%``Higgs radiation off top quarks in e+ e- collisions,''
Phys.\ Lett.\ B {\bf 441}, 383 (1998)
[arXiv:hep-ph/9808433].
%%CITATION = HEP-PH 9808433;%%

%\cite{Asner:2001vh}
\bibitem{Asner:2001vh}
See, for example, D.~Asner {\it et al.},
%``Higgs physics with a gamma gamma collider based on CLIC 1,''
arXiv:hep-ex/0111056.
%%CITATION = HEP-EX 0111056;%%

%\cite{Asner:2001ia}
\bibitem{Asner:2001ia}
D.~M.~Asner, J.~B.~Gronberg, and J.~F.~Gunion,
%``Detecting and studying Higgs bosons in two-photon collisions at a  linear collider,''
arXiv:hep-ph/0110320.
%%CITATION = HEP-PH 0110320;%%
%\cite{Velasco:2002vg}
%\bibitem{Velasco:2002vg}
%M.~M.~Velasco {\it et al.},
%``Photon photon and electron photon colliders with energies below a TeV,''
%in {\it Proc. of the APS/DPF/DPB Summer Study on the Future of Particle Physics 
%(Snowmass 2001) } ed. R.~Davidson and C.~Quigg,
%arXiv:hep-ex/0111055.
%%CITATION = HEP-EX 0111055;%%

\bibitem{muoncollider}
C.~Blochinger {\it et al.},
%``Physics opportunities at mu+ mu- Higgs factories,''
arXiv:hep-ph/0202199.
%%CITATION = HEP-PH 0202199;%%
\end{thebibliography}
\end{document}